\documentstyle[12pt]{article}

\newskip\humongous \humongous=0pt plus 1000pt minus 1000pt

\newif\ifdtup

\def\as{\alpha_S}

\def\abar{{\bar \alpha_S}}

\def\ga{\gamma}
\def\G{{\Gamma}}

\def\ms{$\overline{{\rm MS}}$}

\def\gtap{\raisebox{-.4ex}{\rlap{$\,\sim\,$}} \raisebox{.4ex}{$\,>\,$}}
\def\kper{k_{\perp}}
\def\eav{\langle e^2_f \rangle}
\def\naive{na\"{\i}ve}
\def\np#1#2#3{Nucl.\ Phys.\ B#1 (19#3) #2}
\def\pl#1#2#3{Phys.\ Lett.\ #1B (19#3) #2}
\def\pr#1#2#3{Phys.\ Rev.\ D #1 (19#3) #2}

\def\prl#1#2#3{Phys.\ Rev.\ Lett.\ #1 (19#3) #2}

\def\sj#1#2#3{Sov.\ J.\ Nucl.\ Phys.\ #1 (19#3) #2}
\def\zp#1#2#3{Z.\ Phys.\ C#1 (19#3) #2}

\begin{document}
\begin{titlepage}
\renewcommand{\thefootnote}{\fnsymbol{footnote}}
\begin{flushright}
     DFF 248/4/96\\   April 1996 \\
     hep-ph/9609263
     \end{flushright}
\par \vskip 10mm
\begin{center}
{\Large \bf Physical Anomalous Dimensions at Small $x$ \footnote{Research 
supported in part by EEC Programme {\it Human Capital and Mobility}, Network 
{\it Physics at High
Energy Colliders}, contract CHRX-CT93-0357 (DG 12 COMA).}}
\end{center}
\par \vskip 2mm
\begin{center}
{\bf Stefano Catani}\\  

\vskip 5 mm

{I.N.F.N., Sezione di Firenze}\\
{and Dipartimento
di Fisica, Universit\`a di Firenze}\\
{Largo E. Fermi 2, I-50125 Florence, Italy}
\end{center}

\par \vskip 2mm
\begin{center} {\large \bf Abstract} \end{center}
\begin{quote}

I present a theoretical discussion of the uncertainties related to the QCD
analysis of the proton structure function $F_2(x,Q^2)$ at small $x$. The role
played by the `unphysical' gluon density is pointed out. It is shown how 
the study of more observables can reduce the theoretical uncertainty
and, in particular, an alternative method
of analysis, based on the 
introduction of physical anomalous dimensions, is suggested.

\end{quote}
\vspace*{\fill}
\begin{flushleft}
     DFF 248/4/96\\   April 1996
\end{flushleft}
\end{titlepage}
\renewcommand{\thefootnote}{\fnsymbol{footnote}}

\section{Introduction}\label{intro} 

One of the main outcome of the physics programme carried out at HERA
is the observed striking rise of the proton structure function
$F_{2} (x,Q^2)$ [\ref{HERA}] at small values of the Bjorken variable $x$
($2 \cdot 10^{-5} < x < 10^{-2}$) and high values of the momentum transfer
$Q^2$ ($Q^2 \gtap 2~{\rm GeV}^2$).

The HERA data on $F_2$ represent the first experimental observation of a
cross section increasing faster than logarithmically with the energy (see, for
instance, Ref.~[\ref{C}]). This high-energy behaviour in the hard-scattering
regime is expected if the underlying dynamics is driven by self-interacting 
massless vector bosons, the gluons. Thus, the steep rise of $F_2$ certainly 
confirms one of the basic prediction of perturbative QCD [\ref{first}].

However, the main reason why the HERA data have attracted much theoretical 
attention goes beyond this point. The issue, indeed, is whether the striking 
rise of $F_2$ at small $x$ calls forth a theoretical interpretation in terms
of non-conventional QCD dynamics. In this context, non-conventional QCD
stands for any approach (based either on the original BFKL equation [\ref{BFKL}]
or on $\kper$-factorization [\ref{CCH}-\ref{CH}]) in which the small-$x$
behaviour of $F_2(x,Q^2)$ is studied by resumming
logarithmic corrections of the type $(\as \ln x)^{n}$ to {\em all orders}
in the strong coupling $\as$. By contrast, no small-$x$ resummation
is performed within the conventional QCD (or DGLAP [\ref{first},\ref{AP}]) 
approach: the parton densities of the proton at a fixed input scale $Q_0^2$
are evolved in $Q^2$ according to the Altarelli-Parisi equation evaluated in 
{\em fixed-order} perturbation theory.

The theoretical motivation for the non-conventional approach based on 
resummation is clear. Since multiple gluon radiation in the final state produces
perturbative contributions of the type $(\as \ln x)^{n}$, as soon as
$x$ is sufficiently small (i.e. $\as \ln 1/x \sim 1)$, the fixed-order 
expansion in $\as$ must become inadequate to describe the QCD
dynamics. Thus, in principle, the non-conventional approach is certainly 
more accurate at asymptotically-small values of $x$. The question is whether,
in practice, in the HERA kinematic region we are already approaching this
asymptotic regime.

In my opinion it is quite difficult to answer this question in the context 
of the QCD analysis of the sole $F_2$. Indeed, the small-$x$ rise of $F_2$
can be obtained as the result of two combined effects: the increase of
perturbative scaling violation in the small-$x$ region and the intrinsic
non-perturbative steepness of the gluon density. These perturbative and 
non-perturbative components are mixed up not only on the phenomenological side
but, more importantly, on theoretical basis. Since the QCD description of
a single observable, namely $F_2$, requires the introduction of two 
non-perturbative inputs, quark and gluon densities, the distinction between 
perturbative and non-perturbative components is strongly dependent on their
own definition rather than on the underlying dynamics. 

A better understanding of QCD physics at small $x$ can be achieved by 
considering more observables and thus (over-)constraining the definition of
the parton densities. In particular, by simply using two hadronic observables 
one can formulate the dynamics of scaling violation enterely in terms of 
perturbative quantities that play the role of physical anomalous dimensions.
These anomalous dimensions are unambiguously computable in QCD perturbation 
theory and thus they theoretically appear as golden quantities for comparing
the conventional and non-conventional approaches.

The outline of the paper is as follows. In Sect.~\ref{gluon} I briefly review 
the
theoretical and phenomenological status of the QCD analysis of $F_2(x,Q^2)$.
In particular, I qualitatively discuss the theoretical uncertainties relative
to different perturbative approaches and to the small-$x$ behaviour of the gluon
density. In Sect.~\ref{inva} I introduce the physical anomalous dimensions that
control the $Q^2$ evolution of $F_2(x,Q^2)$ and of the longitudinal structure 
function $F_L(x,Q^2)$. The main features of these anomalous dimensions are 
discussed in Sect.~\ref{pad}, while in Sect.~\ref{bsx} I present their explicit
expressions in resummed perturbation theory at small $x$. Additional 
observations on the relationship between physical anomalous dimensions and 
parton  model are considered in Sect.~\ref{parton}. Section \ref{qqsec} deals 
with physical anomalous dimensions for heavy-flavour structure functions. 
In particular, it points out the kinematical features of the physical 
anomalous dimensions for observables that depend on several large-momentum 
scales. Some general comments
are left to Sect.~\ref{summa}.

\section{The proton structure function $F_2$ and the gluon density}
\label{gluon}

The master equations for the perturbative-QCD study of the proton structure
function at small $x$ are as follows 
\begin{eqnarray} 
F_{2} (x,Q^2) &=&
\eav {\tilde f}_{S} (x,Q^2) + \; \dots \; + {\cal O}(1/Q^2) \;\; ,
\label{F2} \\  \frac{dF_{2} (x,Q^2)}{d\ln Q^2} &=& 
\eav \int^{1}_{x} dz \left[ \,P_{SS}(\as(Q^2), z) \;{\tilde
f}_{S} \left(x/z, Q^2\right)  \right. \nonumber \\ 
 &+& \left. P_{Sg}(\as(Q^2), z) \;{\tilde f}_{g}
\left(x/z,Q^2\right) \right] + \; \dots \; + {\cal O}(1/Q^2)  \;\; ,
\label{dF2} 
\end{eqnarray} 
\begin{eqnarray}\label{dfg}  
\frac{d{\tilde f}_{g} (x,Q^2)}{d \ln
Q^2} &=& \int^{1}_{x} dz \left[  P_{gq}(\as(Q^2), z) \;{\tilde
f}_{S} \left( x/z,Q^2\right)  \right. \nonumber \\ &+& \left.
P_{gg}(\as(Q^2), z) \;{\tilde f}_{g} \left(
x/z,Q^2\right) \right] \;\; .  
\end{eqnarray}
where $e_{f}$ is the electric charge of each
quark with flavour $f, \eav = (\sum^{N_{f}}_{f=1}
e^{2}_{f})/N_{f}$ and $N_{f}$ is the number of active flavours. In
Eqs.~(\ref{F2}-\ref{dfg}) I am using the same notation as in
Ref.~[\ref{CH}]. Thus, the singlet density ${\tilde f}_{S}$ and the gluon
density ${\tilde f}_{g}$ are related to the usual quark (antiquark) and gluon
densities $f_{q_{f}} \;(f_{{\bar q}_f})$ and $f_{g}$ by the following relations
\begin{equation} 
{\tilde f}_{S} (x,Q^2) = x \sum_{f} \left[ f_{q_{f}}
(x,Q^2) + f_{\bar q_{f}} (x,Q^2) \right] \; , \;\;\;{\tilde f}_{g}
(x,Q^2) = x f_{g} (x,Q^2) \;\; , \label{P} 
\end{equation} 
and the
quark splitting function $P_{SS}$ and $P_{Sg}$ are given in terms of the
customary Altarelli-Parisi splitting functions $P_{ab}$ as follows
\begin{equation} 
P_{Sg} (\as, x) = 2 N_{f} P_{q_{i}g} (\as,
x) \; , \;\;\;P_{SS} (\as, x) = \sum_{j}  \;[ P_{q_{i}q_{j}} (\as,
x) +  P_{q_{i}{\bar q_{j}}} (\as, x) ] \;\;. \label{SP} 
\end{equation} 
The dots and the term ${\cal O}(1/Q^2)$ on the right-hand side of
Eqs.~(\ref{F2},\ref{dF2}) respectively denote the flavour non-singlet
component and higher-twist contributions. The contribution of the 
non-singlet component to Eqs.~(\ref{F2},\ref{dF2}) is quantitatively
negligible at small-$x$ and it will be neglected throughout the paper. However,
its inclusion is formally straightforward (see, for instance, Eqs.~(5.3-5.5)
in Ref.~[\ref{CH}]).

The basis for Eqs.~(\ref{F2}-\ref{dfg}) is provided by the factorization theorem
of mass singularities [\ref{CSS}].  According to this theorem the 
({\em perturbatively calculable}) splitting functions $P_{ab}(\as,x)$ and the
({\em phenomenological}) parton densities ${\tilde f}_a(x,Q^2)$ are not 
separately physical observables. Only proper combinations (convolutions) of them
(for instance, the right-hand sides of Eqs.~(\ref{F2},\ref{dF2})) are related
to measurable quantities. Therefore one has some freedom (ambiguity) in defining
splitting functions and parton densities. This freedom is called 
factorization-scheme dependence and follows from the fact that hadron scattering
cross sections cannot be computed within a purely perturbative framework. The
factorization theorem states that at high momentum transfer $Q$, the 
perturbatively non-calculable component of {\em all} the cross sections is 
factorizable in {\em few} universal (process independent) parton distributions.
These parton distributions can be defined using experimental information on
an {\em equal} number of hadronic observables at a certain scale. Having done 
that, the high-$Q^2$ behaviour of all the hadronic cross sections can be
unambiguosly (modulo power suppressed corrections) computed by using 
perturbation theory.
 
Equations (\ref{F2}) and (\ref{dF2}) refer to the so-called DIS factorization 
scheme\footnote{Equation (\ref{dfg}) does not involve (at least, directly) any 
physical observable. Thus, it takes the same form in any scheme. Of course, the
gluon splitting functions $P_{gq}$ and $P_{gg}$ have to be consistently
computed in the corresponding factorization scheme.}[\ref{DIS}]. 
In this scheme, Eq.~(\ref{F2}) actually represents the definition 
of the singlet-quark density ${\tilde f}_{S}$. The true dynamical information is
instead contained in the scaling violations of $F_2$ as described by
Eq.~(\ref{dF2}) and by the analogous evolution equation (\ref{dfg})
for the gluon density.

The Altarelli-Parisi splitting functions entering into
Eqs.~(\ref{dF2},\ref{dfg}) are computable in QCD perturbation theory as
a power series expansion in $\as$: 
\begin{equation}\label{pab} 
P_{ab}(\as, x) = \sum^{\infty}_{n=1} \left( \frac{\as}{2\pi}
\right)^{n} \;\; P_{ab}^{(n-1)}(x) \;\; , \nonumber 
\end{equation} 
and
the coefficients $P_{ab}^{(n-1)}(x)$ in this series can be calculated (at
least, in principle) to any order $n$ in $\as$.

Conceptually, the content of the QCD analysis of $F_2$ according to the master 
equations written above is the following. One first computes the
splitting functions in Eq.~(\ref{pab}) to a given perturbative accuracy.
Then, by using the experimental information on $F_{2}$ and $dF_{2}/d \ln Q^2$,
from Eqs.~(\ref{dF2},\ref{dfg}) one can determine
the quark and gluon densities as functions of $x$ and $Q^2$. Finally,
Eq.~(\ref{dfg}) enters as self-consistency check of the QCD evolution
equations.

In practice, the QCD analysis proceeds as follows. One assigns a certain
parametrization for the parton densities at a given input scale $Q^2_0$.
Then, inserting this parametrization into Eqs.~(\ref{F2}-\ref{dfg}), one can
fit the input parameters to the experimental data.

It is worth emphasizing a point that is independent of the actual procedure
used in the QCD analysis. Whilst the $F_2$ data uniquely determine the 
(DIS scheme) quark density, the measurement of $dF_2/d\ln Q^2$ does not give 
access directly to the determination of the gluon density ${\tilde f}_{g}$, 
but rather to that
of the product (convolution) $P_{Sg} \otimes {\tilde f}_{g}$. Since 
$P_{Sg}$ is evaluated in a certain theoretical framework (that is, at a given
perturbative order or in resummed perturbation theory, in a certain 
factorization scheme and so forth), the ensuing   
${\tilde f}_{g}$ turns out to be `theory-dependent'.

\subsection{Fixed-order perturbation theory}
\label{fo}

Only the first two terms  $P_{ab}^{(0)}(x)$, $P_{ab}^{(1)}(x)$ of the 
perturbative expansion (\ref{pab}) of the splitting functions are exactly known
(i.e. known for any values of $x$) [\ref{CFP}]. In the 
conventional approach these leading order (LO) and next-to-leading order (NLO)
terms are used as theoretical inputs for the
QCD analysis of $F_2$. It turns out that the HERA data can be succesfully
described [\ref{fit}-\ref{DOUBLE}] by parton densities having the following
small-$x$ behaviour
\begin{equation} 
{\tilde f}_{S} (x,Q^2_{0}) \simeq x^{-\lambda_{S}} \;
, \;\; {\tilde f}_{g} (x,Q^2_{0}) \simeq x^{-\lambda_{g}} \label{PB}
\end{equation} 
with $\lambda_{S} \sim \lambda_{g} = 0.2 \div 0.3$
at the input scale $Q^2_{0} \sim$ 4 GeV${}^{2}$.

Up to the second order in $\as$, the quark splitting functions $P_{SS}$,
$P_{Sg}$ in Eq.~(\ref{dF2}) are essentially flat at small $x$, whilst
the gluon splitting functions are steeper and behave as follows\footnote{The
scheme dependence of the splitting functions appears only starting from 
two-loop order. In particular, in two-loop order this dependence is pretty
mild at small $x$.}
\begin{equation}
P_{gg}(\as,x) \simeq \frac{C_A}{C_F} \;P_{gq}(\as,x)  
\simeq \frac{\abar}{x} \;\;,
\end{equation}
where $C_A = N_c, \,C_F = (N_c^2-1)/(2N_c)$, $N_c=3$ is the number of colours
and I have defined $\abar \equiv C_A\as/\pi$. Thus, the phenomenological success
of the NLO QCD approach tells us that the rise of $F_2$ at small $x$
is due to the DGLAP evolution in the gluon channel (i.e. it is 
due to Eq.~(\ref{dfg}))
combined with a steep behaviour $( \,\sim x^{-0.2} \,)$ of the input densities
at $Q^2_{0} \sim$ 4 GeV${}^{2}$.

\subsection{Resummed perturbation theory}
\label{rpt}

The basis for the non-conventional QCD approach is provided by the BFKL 
equation [\ref{BFKL}]. Starting from it, a formalism that is able to combine 
{\em consistently} small-$x$ resummation with the QCD factorization theorem 
has been set up in the last few years. This formalism,
known as $k_{\perp}$-factorization or high-energy factorization, was first
discussed to leading-order accuracy in Refs.~[\ref{CCH}-\ref{CCHLett}]
and then was extended to higher-orders in Refs.~[\ref{CHLett},\ref{CH}].
In the high-energy
factorization approach, one ends up with the usual QCD evolution
equations (namely,  Eqs.~(\ref{F2}-\ref{dfg}) in the case
of the proton structure function $F_{2}$) but the 
splitting functions  $P_{ab}(\as,x)$ in Eq.~(\ref{pab}) (and, in general, 
the process-dependent coefficient functions: see Sects.~\ref{inva}-\ref{qqsec})
are no longer 
evaluated in fixed-order perturbation theory. They are indeed supplemented 
with the all-order resummation of the {\em leading} 
$(\frac{1}{x} \as^{n} \ln^{n-1}x)$, {\em next-to-leading}
$(\frac{1}{x} \as^{n} \ln^{n-2}x)$ and, possibly, subdominant
$(\frac{1}{x} \as^{n} \ln^{m}x, \;m<n-2)$ contributions at small $x$.
Note, also, that this resummation can be performed by having full control
of the factorization-scheme dependence of splitting (and coefficient) functions 
and parton densities [\ref{CCHLett},\ref{CH},\ref{Ciaf}].

The present theoretical status of small-$x$ resummation is the 
following\footnote{I refer to Sects.~\ref{inva} and \ref{qqsec}
for the resummation in the 
process-dependent coefficient functions.}.
The leading-logarithmic (LL) contributions to the gluon splitting functions
$P_{gg}(\as,x)$, $P_{gq}(\as,x)$ are known [\ref{BFKL},\ref{CCHLett},\ref{Jar}].
Their resummation leads to a very steep (power-like) asymptotic behaviour:
\begin{equation}\label{gasym}
P_{gg}(\as,x) \vert_{\rm {asym.}} \simeq \frac{C_{A}}{C_{F}}
P_{gq}(\as,x) \vert_{\rm {asym.}} \sim \abar \;x^{-(1+\lambda_L)} \;\;,
\end{equation} 
where the power $1+\lambda_L = 1 + 4 \abar \ln 2 \simeq 1 + 2.65 \as$ is 
the so-called
intercept of the perturbative QCD pomeron. The complete
next-to-leading logarithmic (NLL)
contributions to the gluon splitting functions are not yet known and 
calculations are in progress [\ref{FLF},\ref{DD},\ref{Cam}]. In particular,
the contributions proportional to $N_f$ in $P_{gg}$ have been evaluated 
recently [\ref{Cam}]. Owing to the gluon dominance
at high energy, the quark splitting functions $P_{Sg}(\as,x)$, $P_{SS}(\as,x)$
do not contain LL contributions. However, the NLL
terms are completely known [\ref{CHLett},\ref{CH}] to all orders in
perturbation theory.

Having developed a resummed perturbative expansion to the same (modulo the
still unknown NLL 
terms in the gluon sector) degree of theoretical 
accuracy as the fixed-order perturbative expansion, one can set up
a fully consistent non-conventional QCD approach [\ref{CH}]. This is 
accomplished [\ref{EHW},\ref{BFres}] by
adding leading and next-to-leading logs to one- and two-loop contributions
(after subtracting the resummed logarithmic terms, in order to avoid double 
counting) in the splitting functions $P^{ab}(\as,x)$, thus obtaining a 
perturbative framework which is everywhere at least as good as the
fixed-order expansion, and much better as $x$ becomes small.

After the first numerical analyses [\ref{AKMS}] within the 
$\kper$-factorization framework,
phenomenological studies based on resummed perturbation theory have 
been performed during the last year [\ref{EHW},\ref{BFres},\ref{FRT}]. They
have shown that, likewise the conventional QCD analysis,
the non-conventional approach can accomodate the parton densities to
provide a description of the HERA data on $F_2$.

The \naive\ explanation for that could be that the inclusion of the 
resummed logarithmic corrections produces a small effect in the 
Altarelli-Parisi splitting functions. Actually, this is not the case.

Indeed, it is true that the resummation of the leading terms 
$\frac{1}{x} \as^{n} \ln^{n-1} x$ in the gluon splitting functions 
has a moderate impact on the scaling violations of $F_{2}$
in the kinematical range presently investigated at HERA. The situation 
is however different in the quark channel, that is, in the evolution equation
(\ref{dF2}). The measured large value of $dF_{2}(x,Q^2)/d\ln Q^2$ 
at small $x$ calls for a quite steep product
(convolution) $P_{Sg} \otimes {\tilde f}_{g}$. In the conventional
(fixed-order) perturbative analysis this condition  can be fulfilled only
by choosing a quite steep input distribution ${\tilde f}_{g}$. After 
resummation of the next-to-leading terms $\frac{1}{x} \as^{n} \ln^{n-2}x$, 
the quark splitting functions $P_{Sg}(\as,x)$ and $P_{SS}(\as,x)$ are 
much steeper than the corresponding splitting functions evaluated in two-loop
order\footnote{I refer to [\ref{SDIS}] for a more detailed 
discussion on the small-$x$ behaviour of the resummed splitting functions.}.
Thus, the non-conventional approach can succeed in describing the small-$x$ rise
of $F_2$ by using parton densities that at the input scale $Q_0^2$ are less
steep than those needed in the fixed-order approach.

From this result one may conclude that the conventional and non-conventional 
approaches are phenomenologically equivalent: the HERA data on $F_2$ at 
sufficiently high $Q^2$ cannot distinguish steep input densities from steep 
dynamical evolution. 

The conclusion is instead different from a theoretical viewpoint.  Since the 
resummation of the NLL contributions leads to a large effect on $F_2$, there 
is no justification for truncating the QCD perturbative expansion at NLO: the
fixed-order expansion approach is thus theoretically disfavoured. The only 
caveat against such a firm conclusion is that the NLL terms in the 
gluon sector are still unknown: they may lead to a large and opposite effect
with respect to those in the quark channel.

At the same time, since the known NLL contributions produce large corrections 
on $F_2$, one can expect that subleading terms may still have a sizeable effect.
Thus, at present, perturbative QCD predictions for the small-$x$ behaviour of 
$F_2$ suffer from substantial theoretical uncertainties [\ref{EHW}]. A better 
understanding of subleading contributions is necessary to reduce these 
uncertainties [\ref{EHW},\ref{subeffects}].

The QCD analysis of the sole proton structure function $F_2$, moreover, is
affected by an even larger indeterminacy related to the difficulty 
in disentangling perturbative and non-perturbative effects. In order to 
clarify this point, let me briefly consider the issue of the 
factorization-scheme dependence [\ref{Ciaf},\ref{SDIS},\ref{BFscheme}].

\subsection{Factorization-scheme dependence}
\label{fsd}

As discussed in the first part of this Section, the parton densities are not
physical observables and, in particular, they are not calculable in perturbation
theory. In the perturbative framework they are defined apart from an overall 
perturbative function. Thus, starting from the DIS factorization scheme
considered so far, one can introduce a new factorization scheme of 
DIS type\footnote{More general factorization schemes, like for instance the 
${\overline {\rm MS}}$ scheme, are considered in 
Refs.~[\ref{CH},\ref{BFscheme}].} 
(i.e. a scheme in which the physical identification of the quark density with 
the proton structure function as in Eq.~(\ref{F2}) remains valid) by defining
a new gluon density ${\tilde f}_g^{(new)}$ as follows [\ref{SDIS}]
\begin{equation}\label{gnew}
\!{\tilde f}_g^{(new)}(x,Q^2) = 
{\tilde f}_g(x,Q^2) + \int_x^1 \frac{dz}{z} \left[ u(\as(Q^2),z) 
\,{\tilde f}_g(x/z,Q^2) + \frac{C_F}{C_A} \;v(\as(Q^2),z)  
\,{\tilde f}_S(x/z,Q^2) \right] .
\end{equation}
Here $u(\as,z)$ and $v(\as,z)$ are functions that can be expanded as power 
series in $\as$ and vanish for $\as=0$. As for their functional dependence on
$z$, it is
quite arbitrary. The only constraints are that $u(\as,z)$ and $v(\as,z)$ contain
at most NLL terms of the type $\as(\as \ln z)^n$ for $z \to 0$ and that these 
NLL terms are equal in $u$ and $v$.

The dynamical evolution equations (\ref{dF2}) and (\ref{dfg}) can be written in 
terms of the new gluon density (\ref{gnew}) and of new Altarelli-Parisi
splitting functions. The above constraints guarantee that the new splitting
functions have the same LL behaviour as the DIS-scheme splitting functions,
so that the dominant perturbative dynamics is left unchanged. 

The freedom of arbitrarily choosing the factorization scheme is not a particular
feature of small-$x$ dynamics. The transformation in Eq.~(\ref{gnew}) can
be applied in the small-$x$ as well as in the large-$x$ regions. In general, its
effect amounts to a redefinition of the input parton densities that is 
perturbatively under control. The effect, instead, can be quite large in the
small-$x$ region because each power of $\as$ can be accompanied by an enhancing
logarithmic factor of $\ln 1/x$.

In order to quantify the theoretical uncertainty related to the scheme 
dependence, let us consider the simplest case in which the splitting functions 
in Eqs.~(\ref{dF2},\ref{dfg}) are evaluated to LL accuracy. Thus, we can perform
the scheme transformation in Eq.~(\ref{gnew}) by choosing any NLL functions
$u$ and $v$ and, in particular, we can set $u(\as,z) = v(\as,z) = A 
\as z^{-K\as}$, where $A$ and $K$ are constants of order unity. Assuming the 
extreme case of flat input densities, this leads to the following
factorization-scheme uncertainty 
\begin{equation}
\label{delf}
\delta {\tilde f}_g(x) = {\tilde f}_g^{(new)}(x) - {\tilde f}_g(x)
= \frac{A}{K} \,(x^{-K\as} - 1) \sim \frac{A}{K} \,x^{-K\as} \;\;.
\end{equation}
This implies that, from the
QCD analysis of $F_2$ to LL accuracy, one cannot argue whether steep input 
densities have a non-perturbative origin or rather mimic higher-order
perturbative effects. Of course, using the NLL expressions for the splitting 
functions one reduces the factorization-scheme uncertainty by a factor of $\as$.
For the case considered above one obtains  
$\delta {\tilde f}_g(x) \sim \as \,x^{-K\as}$ that, however, still represents
a substantial indeterminacy.

In order to gain more theoretical accuracy one should compute higher
perturbative orders in the Altarelli-Parisi splitting functions. The same goal
can be achieved in a simpler manner by eliminating the factorization-scheme
uncertainty, that is, by relating the gluon density to other physical 
observables. 

Actually, there is one more reason for studying the small-$x$ behaviour of
physical observables other than $F_2$. As a matter of fact, the large effect
on $F_2$ of the NLL contributions in the quark channel might be, in a sense,
spurious or, more precisely, related to the use of certain factorization schemes 
[\ref{Ciaf},\ref{SDIS}].

For instance (and quite strikingly), one can choose the functions
$u(\as,z)$ and $v(\as,z)$ in such a
way that all the NLL terms in the quark channel are removed from the (new)
quark splitting functions $P_{Sg}$, $P_{SS}$ and absorbed into the redefinition
(\ref{gnew}) of the gluon density [\ref{SDIS}]. The only price one has to pay
consists in the introduction of additional NLL terms in the gluon splitting 
functions. It turns out that the effect of these additional terms is 
quantitatively small [\ref{BFres}]. Therefore, within this factorization scheme
(called SDIS scheme in Ref.~[\ref{SDIS}]) the non-conventional 
approach
to NLL accuracy and the conventional approach to NLO are, in practice,
indistinguishable  
(apart from the caveat on the unknown NLL terms in the gluon channel)
as for the QCD analysis of $F_2$. The introduction of the SDIS scheme thus 
provides a more formal argument to explain the phenomenological equivalence 
of the conventional and non-conventional approaches that has been pointed out
in Sect.~\ref{rpt}.

This equivalence may appear as due to a particular algebraic trick or to
fine-tuning of the factorization scheme with no physical content. Actually, 
this is not necessarely
the case. The factorization-scheme dependence, rather than an ambiguity in 
higher-order perturbative coefficients, has to be regarded more 
physically as a parametrization of our ignorance in factorizing 
perturbative from non-perturbative physics.
At present, the proton structure function $F_2$ (i.e. Eqs.~(\ref{F2}-\ref{dfg}))
provides experimental/theoretical information that is not sufficiently accurate
to disentangle perturbative from non-perturbative dynamics at small-$x$.

In order to have better control on the perturbative dynamics, one should 
consider the small-$x$ behaviour of other physical observables. Indeed, by 
means of the transformation from the DIS to the SDIS schemes (almost) no
trace of small-$x$ perturbative contributions is left in $F_2$ and all
the resummation effects are moved to other physical quantities. These effects 
can be sizeable. It may also happen that the resummed contributions are almost 
universal, in the sense that, in the kinematic regions that are experimentally 
accessible, they produce very similar quantitative effects in all physical 
observables. In this case, these contributions can be consistently absorbed
into the non-perturbative parton densities and fixed-order perturbation theory
can be safely used throughout.

\section{Factorization-theorem invariants at small $x$}
\label{inva}

The theoretical motivations for studying the small-$x$ behaviour of
several different physical observables have been pointed out in the previous
Section. As discussed in Refs.~[\ref{CCH},\ref{CAMICI}] and furtherly
elaborated on in Ref.~[\ref{Cam}], 
this study can be
performed by considering properly defined $K$-factors (ratios of hadronic
cross sections), which are factorization-scheme independent. Analogously, one 
can introduce factorization-scheme invariants that relate the scaling 
violations of different structure functions. These invariants are discussed
in the rest of this paper.

Among the observables that one can consider, the longitudinal structure function 
$F_{L}$ of the proton is becoming increasingly topical. On the 
experimental 
side, data on $F_L(x,Q^2)$ at small $x$ will be available soon from HERA. On 
the theoretical side,
this quantity is known to a sufficient accuracy. 

In order to make more explicit this statement about the theoretical accuracy 
of $F_L$, let me recall that, using the factorization theorem of mass 
singularities, $F_L$ is given as follows
\begin{eqnarray} \label{FL}
F_{L} (x,Q^2) &=& \eav \int^{1}_{x} \frac{dz}{z} 
\left[ \, C_{L}^S(\as(Q^2),z)   
\;{\tilde f}_{S} \left(x/z, Q^2 \right) \right. \nonumber \\ 
&+& \left. C_L^g(\as(Q^2),z)  
\;{\tilde f}_{g}\left(x/z,Q^2 \right) \right] +
 \; \dots \; + {\cal O}(1/Q^2)  \;\; ,
\end{eqnarray} 
where ${\tilde f}_{S}$ and ${\tilde f}_{g}$ are the same parton densities
that enter into Eqs.~(\ref{F2}-\ref{dfg}) and, as in Eqs.~(\ref{dF2},\ref{dfg}),
the dots and the term ${\cal O}(1/Q^2)$ respectively denote flavour 
non-singlet and higher-twist contributions. In any given factorization scheme
the coefficient functions $C_L^S$ and $C_L^g$ in Eq.~(\ref{FL}) are computable
in QCD perturbation theory according to the following power series expansion
\begin{equation}\label{cpert} 
C_L^a(\as,x) = \sum_{n=1}^{+\infty} 
\left( \frac{\as}{2\pi} \right)^n \;C_L^{a \,(n-1)}(x) \;\;.
\end{equation}
Both the LO and NLO coefficients $C_L^{a \,(0)}(x), \,C_L^{a \,(1)}(x)$ have 
been computed for any value of $x$ [\ref{LCF}]. Correspondingly, in resummed 
perturbation theory all the NLL terms $\ln^{n-1}x$ in $C_L^{a \,(n)}(x)$ 
are known [\ref{CH}]. Owing to this theoretical information, 
Eqs.~(\ref{F2}-\ref{dfg}) can be supplemented with Eq.~(\ref{FL}) thus 
eliminating the factorization scheme uncertainty. To this purpose one should
introduce the parton densities ${\tilde f}_{S}$ and ${\tilde f}_{g}$ extracted
from Eqs.~(\ref{F2}-\ref{dfg}) into Eq.~(\ref{FL}). Theoretical consistency
simply requires that splitting functions and coefficient functions are 
evaluated to the corresponding accuracy, that is, to NLO in the conventional
approach and including NLL terms in resummed perturbation theory.

\subsection{Physical anomalous dimensions}
\label{pad}

The unphysical role played by the parton densities within this context is clear.
Indeed, one can write down evolution equations that involve only physical
observables and perturbative quantities.
Starting from Eqs.~(\ref{F2},\ref{FL}) and performing straightforward
algebraic manipulations, one first express the parton densities ${\tilde f}_S$
and ${\tilde f}_g$ as functions of $F_2$ and $F_L$. Then, inserting the 
expressions derived in this manner into Eqs.~(\ref{dF2},\ref{dfg}), one obtains
the following dynamical equations
\begin{eqnarray} 
\frac{dF_{2} (x,Q^2)}{d\ln Q^2} &=& \int^{1}_{x} \frac{dz}{z} \left[ 
\,\G_{22}(\as(Q^2), z) \;
F_{2} \left(x/z, Q^2\right)  \right. 
%\nonumber \\ &+& 
+ \left. \G_{2L}(\as(Q^2), z) \;F_{L}\left(x/z,Q^2\right) 
\right] \nonumber \\
&+& \; \dots \; + {\cal O}(1/Q^2)  \;\; ,
\label{dF2p} \\
\frac{dF_{L} (x,Q^2)}{d\ln Q^2} &=& \int^{1}_{x} \frac{dz}{z} \left[ 
\,\G_{L2}(\as(Q^2), z) \;
F_{2} \left(x/z, Q^2\right)  \right. 
%\nonumber \\ &+& 
+ \left. \G_{LL}(\as(Q^2), z) \;F_{L}\left(x/z,Q^2\right) 
\right] \nonumber \\ 
&+& \; \dots \; + {\cal O}(1/Q^2)  \;\; ,
\label{dFLp} 
\end{eqnarray} 

From a formal viewpoint Eqs.~(\ref{dF2p},\ref{dFLp}) may appear equivalent to
Eqs.~(\ref{dF2},\ref{dfg}). However, Eqs.~(\ref{dF2p},\ref{dFLp}) relate the
scaling violations of two physical observables, namely $F_2$ and $F_L$,
to the actual value of the same observables. It follows that the
kernels $\G_{ij}(\as(Q^2), x)$ (with $i,j=2,L$) are {\em physical observables}
as well. Owing to the formal resemblance to the Altarelli-Parisi splitting 
functions, the kernels $\G_{ij}(\as(Q^2), x)$ can be considered as 
{\em physical splitting functions}.

The main physical properties of the kernels $\G_{ij}(\as(Q^2), x)$ are that 
$i)$ each of them is consistently computable in QCD perturbation theory (modulo
higher-twist corrections that are suppressed by some power of $1/Q$ in the 
hard-scattering regime) and $ii)$ each of them is a factorization-theorem 
invariant, i.e. it does not depend on both the factorization scheme and
the factorization scale. In other words, from the
viewpoint of perturbative QCD, each $\G_{ij}(\as(Q^2), x)$ is completely
analogous to the celebrated ratio 
\begin{equation}
R_{e^+e^-} = 
\frac{\sigma(e^+e^- \to {\rm hadrons})}{\sigma(e^+e^- \to \mu^+\mu^-)} \;\;,
\end{equation}
%$R_{e^+e^-} = \sigma(e^+e^- \to {\rm hadrons})/\sigma(e^+e^- \to \mu^+\mu^-)$ 
in $e^+e^-$ annihilation.

The perturbative expansions of the physical splitting functions are the 
following
\begin{equation}\label{gllp}
\G_{LL}(\as,x) = 
                   \sum_{n=1}^{+\infty} \left( \frac{\as}{2\pi} \right)^{n} 
                   \;\G_{LL}^{(n-1)}(x)
  = \frac{\as}{2\pi} \left[ \G_{LL}^{(0)}(x) 
                    + \frac{\as}{2\pi} \;\G_{LL}^{(1)}(x) + \dots \right] \;\;,
\end{equation}
\begin{equation}\label{gl2p}
\G_{L2}(\as,x) = 
                   \sum_{n=1}^{+\infty} \left( \frac{\as}{2\pi} \right)^{n+1} 
                   \;\G_{L2}^{(n-1)}(x)
  = \left( \frac{\as}{2\pi} \right)^2 \left[ \G_{L2}^{(0)}(x) 
                    + \frac{\as}{2\pi} \;\G_{L2}^{(1)}(x) + \dots \right] \;\;,
\end{equation}
\begin{equation}\label{g2lp}
\G_{2L}(\as,x) = 
                   \sum_{n=1}^{+\infty} \left( \frac{\as}{2\pi} \right)^{n-1} 
                   \;\G_{2L}^{(n-1)}(x)
  =                \left[ \G_{2L}^{(0)}(x) 
                    + \frac{\as}{2\pi} \;\G_{2L}^{(1)}(x) + \dots \right] \;\;,
\end{equation}
\begin{equation}\label{g22p}
\G_{22}(\as,x) = 
                   \sum_{n=1}^{+\infty} \left( \frac{\as}{2\pi} \right)^{n} 
                   \;\G_{22}^{(n-1)}(x)
  = \frac{\as}{2\pi} \left[ \G_{22}^{(0)}(x) 
                    + \frac{\as}{2\pi} \;\G_{22}^{(1)}(x) + \dots \right] \;\;.
\end{equation}
Note that the expansions for the diagonal kernels $\G_{LL}$,
$\G_{22}$ are completely analogous to that in Eq.~(\ref{pab}) for the 
Altarelli-Parisi splitting functions. The mismatch in the overall power
of $\as$ between the expansions for the diagonal and non-diagonal ($\G_{L2}$,
$\G_{2L}$) kernels is due to the fact that, from a perturbative viewpoint
(or, equivalently, because of the validity of the Callan-Gross relation, 
$F_L=0$, in the \naive\ parton model), 
the longitudinal structure function has to be considered as a physical quantity
of relative order $\as$ with respect to $F_2$, i.e. $F_L  \sim \as F_2$.
Taking this into account, a conventional QCD calculation should consistently
consider the contributions $\G_{ij}^{(0)}(x)$ in Eqs.~(\ref{gllp}-\ref{g22p}) 
as lowest-order terms, $\G_{ij}^{(1)}(x)$ as next-order terms and so forth.
                    
Obviously, since the kernels $\G_{ij}$ are physical observables, they are
renormalization-group invariant quantities. It follows that, if
computed in fixed-order perturbation theory they should exibits the customary 
dependence on the renormalization scale $\mu$. Thus, to be more precise,
in the evolution equations (\ref{dF2p},\ref{dFLp}) one has to perform the
replacement $\G_{ij}(\as(Q^2),x) \to \G_{ij}(\as(\mu^2),Q^2/\mu^2,x)$.
Equations (\ref{gllp}-\ref{g22p}) refer to the perturbative expansion
of $\G_{ij}$ for $\mu=Q^2$. In general one obtains:
\begin{equation}\label{gijpmu}
\!\G_{ij}\!\left(\as(\mu^2),\frac{Q^2}{\mu^2},x\right) = 
  \left( \frac{\as(\mu^2)}{2\pi} \right)^p
  \; \left[ \G_{ij}^{(0)}(x) 
                    + \frac{\as(\mu^2)}{2\pi} 
  \left( \G_{ij}^{(1)}(x) - p \;\G_{ij}^{(0)}(x) \,2\pi \beta_0 
  \ln \frac{Q^2}{\mu^2}
  \right) 
    + .. \right] ,
\end{equation}
where $12 \pi \beta_0 = 11 C_A - 2 N_f$ is the first coefficient of the QCD
$\beta$-function and $p=1$ for $\G_{LL}$ and $\G_{22}$, $p=0$ for $\G_{2L}$, 
$p=2$ for $\G_{L2}$.

This discussion on the perturbative features of the kernels $\G_{ij}(\as,x)$
can be summarised by saying that they are infrared and collinear safe 
quantities. Thus, as in the case of the ratio $R_{e^+e^-}$, the $x$-dependent
perturbative coefficients $\G_{ij}^{(n)}$  are computable by first principles
starting from parton-level Feynman diagrams and without carrying out any
factorization procedure  of mass singularities. Nonetheless, since higher-order
perturbative calculations for Altarelli-Parisi splitting functions and 
process-dependent coefficient functions are already available, it is more 
convenient to relate directly the $\G_{ij}^{(n)}$'s to these quantities.

To the purpose of simplifying the notation it is also useful to introduce
the $N$-moments. For any function $g(x)$, I define its $N$-moments $g_N$ in
the usual way:
\begin{equation}
g_{N} \equiv \int^{1}_{0} dx \; x^{N-1} \;g(x) \;\;.
\end{equation}
Thus, for instance, the evolution equations (\ref{dF2},\ref{dfg}) become:
\begin{equation} 
\frac{dF_{2,N} (Q^2)}{d\ln Q^2} = \;
\eav  \left[ \,\ga_{SS, \,N}(\as(Q^2)) \;{\tilde
f}_{S, \,N}(Q^2) + \ga_{Sg, \,N}(\as(Q^2)) 
\;{\tilde f}_{g, \,N}(Q^2) \right] \;\; , 
\end{equation} 
\begin{equation}
\label{dfgn}
\frac{d{\tilde f}_{g, \,N}(Q^2)}{d \ln Q^2} = 
 \ga_{gq, \,N}(\as(Q^2)) 
\;{\tilde f}_{S, \,N}(Q^2) +
\ga_{gg, \,N}(\as(Q^2)) \;{\tilde f}_{g, \,N}(Q^2)  \;\;,  
\end{equation}
where the {\em anomalous dimensions} $\ga_{ab, \,N}(\as)$ are related to the 
$N+1$-moments of the Altarelli-Parisi splitting functions, that is,
\begin{equation}
\gamma_{ab,N}(\as) \equiv \int^{1}_{0} dx \; x^{N}
P_{ab}(\as, x) = \;P_{ab, \,N+1}(\as) \;\; .
\label{AD}
\end{equation}

Analogously, the dynamical equations (\ref{dF2p},\ref{dFLp}) can be rewritten
as follows
\begin{equation}\label{dF2pn} 
\frac{dF_{2,N}(Q^2)}{d\ln Q^2} = 
\,\G_{22, \,N}(\as(Q^2)) \;
F_{2, \,N}(Q^2) + \G_{2L, \,N}(\as(Q^2)) 
\;F_{L, \,N}(Q^2)  \;\; , 
\end{equation} 
\begin{equation}\label{dFLpn} 
\frac{dF_{L, \,N}(Q^2)}{d \ln Q^2} = 
\G_{L2, \,N}(\as(Q^2)) 
\;F_{2, \,N}(Q^2) +
\G_{LL, \,N}(\as(Q^2)) \;F_{L, \,N}(Q^2) \;\;,  
\end{equation}
where $\G_{ij, \,N}(\as)$ are the {\em physical anomalous dimensions},
i.e. the $N$-moments of the physical splitting functions $\G_{ij}(\as,x)$.

The physical anomalous dimensions are related to $\ga_{ab, \,N}$ and to the 
longitudinal coefficient functions in Eq.~(\ref{FL}) by the following equations
\begin{equation}\label{gll}
\G_{LL, \,N} = \left[ \ga_{gg, \,N} 
                    + \frac{C_{L, \,N}^S}{C_{L, \,N}^g} \;\ga_{Sg, \,N}
                    + \frac{d\ln C_{L, \,N}^g}{d\ln Q^2} 
                    \right]_{{\rm DIS}} \;\;,
\end{equation}
\begin{eqnarray}\label{gl2}
\G_{L2, \,N} &=& \left[ C_{L, \,N}^g \ga_{gq, \,N} - C_{L, \,N}^S \ga_{gg, \,N}
            + C_{L, \,N}^S  \left( \ga_{SS, \,N} -
            \frac{C_{L, \,N}^S}{C_{L, \,N}^g} \;\ga_{Sg, \,N} \right)
             \right. \nonumber \\
               &~& \;\; \left.                       
                    + C_{L, \,N}^S \left( \frac{d\ln C_{L, \,N}^S}{d\ln Q^2}
                    - \frac{d\ln C_{L, \,N}^g}{d\ln Q^2} \right)
                    \right]_{{\rm DIS}} \;\;,
\end{eqnarray}
\begin{equation}\label{g2l}
\G_{2L, \,N} = \left[ \frac{\ga_{Sg, \,N}}{C_{L, \,N}^g} 
                     \right]_{{\rm DIS}} \;\;,
\end{equation}
\begin{equation}\label{g22}
\G_{22, \,N} = \left[ \ga_{SS, \,N} 
                    - \frac{C_{L, \,N}^S}{C_{L, \,N}^g} \ga_{Sg, \,N}
                    \right]_{{\rm DIS}} \;\;,
\end{equation}
where I have used the shorthand notation
$\G_{ij, \,N} = \G_{ij, \,N}(\as(Q^2))$,
$\ga_{ab, \,N} = \ga_{ab, \,N}(\as(Q^2))$,
$C_{i, \,N}^a = C_{i, \,N}^a(\as(Q^2))$. In 
Eqs.~(\ref{gll}-\ref{g22})  the subscript DIS on the right-hand side means 
that the quantities inside the square brackets have to be evaluated in the DIS
factorization scheme. Obviously, this does not mean that $\G_{ij, \,N}$ are
scheme dependent. The only point is that their expressions in terms of
$\ga_{ab, \,N}$ and $C_{L, \,N}^a$ are more cumbersome if $\ga_{ab, \,N}$ and 
$C_{L, \,N}^a$ are given in a different factorization scheme.

Both the Altarelli-Parisi splitting functions and the longitudinal coefficient 
functions [\ref{LCF}] are known up to two-loop order. Therefore, using
Eqs.~(\ref{gll}-\ref{g22}), one can obtain the two lowest-order terms 
$\G_{ij}^{(0)}$, $\G_{ij}^{(1)}$ of the physical anomalous dimensions.

\subsection{Behaviour at small $x$}
\label{bsx}

Let me now consider the small-$x$ behaviour of the physical anomalous 
dimensions.
From power-counting arguments, it follows that the most singular terms in the 
perturbative
coefficients $\G_{ij}^{(n)}(x)$ behave as $\G_{ij}^{(n)}(x) \sim 
x\,P^{(n)}(x) \sim (\ln x)^n$ or, equivalently, 
$\G_{ij, \,N}^{(n)} \sim (1/N)^{n+1}$ in $N$-moment space\footnote{
Note that logarithmic contributions of the type $\ln^{n-1}x$
in $x$-space correspond to multiple poles $(1/N)^{n}$ in $N$-space.}.
As in the case of 
the Altarelli-Parisi splitting functions, one expects two entries in the matrix 
of the physical splitting functions which contain leading logarithms.
These two entries are those more directly related to the gluon channel and,
hence, they appear in the evolution equation (\ref{dFLp})  for the longitudinal
structure function. Thus, we have $\G_{LL}^{(n)}(x) \sim \G_{L2}^{(n)}(x) 
\sim (\ln x)^n$. The evolution equation (\ref{dF2p}) is instead more related
to the quark dynamics and thus the corresponding anomalous dimensions
contain only NLL terms, i.e. 
$\G_{2L}^{(n)}(x) \sim \G_{22}^{(n)}(x) \sim (\ln x)^{n -1}$.

As in the case of the fixed-order perturbative expansions, 
Eqs.~(\ref{gll}-\ref{g22}) can be used to obtain resummed logarithmic 
expressions at small $x$ for the physical anomalous dimensions. The resummation
programme carried out in Refs.~[\ref{CCH},\ref{CH}] leads to analytic formulae 
given in terms of the LL contributions to the gluon anomalous
dimensions, that is,
\begin{equation}
\gamma_{gg, \,N}(\as) = \gamma_{N}(\as) + {\cal O}\left( \as (\as/N)^{n} \right) 
\; . 
\label{gad}
\end{equation}
Here, $\gamma_{N}(\as)$ is the BFKL anomalous dimension [\ref{BFKL},\ref{Jar}]
and is obtained by solving the implicit equation 
\begin{equation}
1 = \frac{{\bar \as}}{N} \;\chi
\left(\gamma_{N}(\as)\right) \; , 
\label{LAD}
\end{equation}
where the characteristic function $\chi(\gamma)$ is expressed in terms of
the Euler $\psi$-function as follows
\begin{equation}
\chi(\gamma) = 2\psi(1) - \psi(\gamma) - \psi(1-\gamma) \; .
\label{CHI}
\end{equation}
Having recalled these results,
I am now in a position of presenting all-order resummed formulae 
for the physical anomalous dimensions, starting from the leading components 
$\G_{LL}$ and $\G_{L2}$. 

Since quark splitting functions and 
longitudinal coefficient functions are subleading at small $x$, Eqs.~(\ref{gll})
and (\ref{gad}) immediately gives
\begin{equation}\label{gllr}
\G_{LL, \,N}(\as) = \ga_N(\as) \;
                    + {\cal O}\left( \as(\as/N)^k \right)  \;\;.
\end{equation}
Note also that, using the known next-to-leading results for $\ga_{Sg}$ and
$C_L^a$, as soon as the next-to-leading contributions to $\ga_{gg}$ will
be evaluated, one can provide the full NLL corrections to $\G_{LL}$ [\ref{Cam}].

Incidentally, Eq.~(\ref{gllr}) clearly shows that the BFKL anomalous dimensions
$\ga_N(\as)$, being related to the small-$x$ behaviour of 
$\G_{LL}$, is a physical quantity. On the contrary, the gluon anomalous 
dimensions $\ga_{gg, \,N}$ are factorization-scheme dependent and, in general, 
one might expect that this scheme-dependence affects also their LL behaviour.
  
As can be seen from Eqs.~(\ref{gl2},\ref{g2l},\ref{g22}), to the purpose of
evaluating the LL terms in $\G_{L2}$, as well as the next-to-leading 
terms in $\G_{2L}$ and $\G_{22}$, it is {\em not} sufficient to know the leading
contributions in the gluon channel, i.e. in $\ga_{gq}$ and $\ga_{gg}$. One has 
to use the full information provided by the next-to-leading order resummation
performed in Ref.~[\ref{CH}]. This feature emphasizes once more that the 
standard
anomalous dimensions and coefficient functions are not physical observables.
The power counting of small-$x$ logarithms is different for physical observables
and leading and next-to-leading logarithms in anomalous dimensions and 
coefficient functions get (slightly) mixed up. Without the NLL
calculations in Ref.~[\ref{CH}], no LL analysis of physical 
quantities at small $x$ can be carried out. 

Using the results for $\ga_{gq}$, $\ga_{Sg}$, $C_L^a$ obtained in 
Ref.~[\ref{CH}], Eq.~(\ref{gl2}) gives the following expression for the LL
terms in $\G_{L2}$
\begin{equation}\label{gl2r}
\G_{L2, \,N}(\as) = \frac{\as}{2\pi} \left\{ \left( \frac{C_F}{C_A} 
           \;C_{L, \,N}^{g \,(0)} - C_{L, \,N}^{S \,(0)} \right)
           \;\ga_N(\as) +\;
           {\cal O}\left( \as(\as/N)^k \right) \right\} \;\;, 
\end{equation}
where $C_{L, \,N}^{a \,(0)}$ are the $N$-moments of the lowest-order coefficient
functions $C_{L}^{a \,(0)}(x)$ in Eq.~(\ref{cpert}). 
One can see that, eventually, also the small-$x$ resummation in $\G_{L2}$ turns 
out to be proportional to the BFKL anomalous dimension. As for the NLL
terms in $\G_{L2}$, part of them can be obtained from those 
known [\ref{CH}] for quark anomalous dimensions and coefficient functions. The
remaining terms require the evaluation of the gluon anomalous dimensions
$\ga_{gg}$ and $\ga_{gq}$ to NLL order {\em and} the computation
of the DIS-scheme coefficient functions $C_{L}^{a}$ to 
{\em next-to-next-to-leading} logarithmic (NNLL) accuracy! This feature is 
consistent with the 
factorization-scheme dependence of Eqs.~(\ref{F2}-\ref{dfg}). Indeed, it is 
straightforward to check that, after having fixed the factorization scheme
to NLL accuracy in Eqs.~(\ref{F2},\ref{dF2}), only the NLL contributions of
$P_{gg}$ in Eq.~(\ref{dfg}) are unambiguously defined: by properly choosing
the scheme transformation in Eq.~(\ref{gnew}), one still has the freedom of
arbitrarily defining the NLL terms in the non-diagonal gluon splitting
function $P_{gq}$. In other words,
the sole calculation of the still unknown gluon 
anomalous dimensions to NLL order will not be
sufficient to provide a consistent theoretical framework for the analysis
of physical observables to NLL accuracy in resummed perturbation theory.

The evaluation of the next-to-leading contributions to the physical anomalous
dimensions $\G_{2L}$, $\G_{22}$ enterely relies on the calculations of the quark
anomalous dimensions in Ref.~[\ref{CHLett}] and of the longitudinal coefficient 
functions in Ref.~[\ref{CH}]. Using these results and 
Eqs.~(\ref{g2l},\ref{g22}), one obtains\footnote{Note that, consistently with 
the logarithmic accuracy of the right-hand sides of Eqs.~(\ref{gl2r},\ref{g22r}),
the $N$-moments of the lowest-order anomalous dimensions and coefficient 
functions can be replaced with their values at $N=0$, namely
$\ga_{SS, \,N=0}^{(0)} = 0 ,\;\; \ga_{Sg, \,N=0}^{(0)} = 
C_{L, \,N=0}^{g \,(0)} = \frac{4}{3} T_R N_f , \;\; 
C_{L, \,N=0}^{S \,(0)} = C_F$.}
\begin{equation}\label{g2lr}
%\G_{2L, \,N}(\as) = 
%          \left[ \frac{1}{1-\ga_N(\as)} + \frac{3}{2} \;\ga_N(\as) \right] \;
%          \left[ 1 + {\cal O}\left( \as(\as/N)^k \right) \right] 
\frac{\as}{2\pi} \;\G_{2L, \,N}(\as) = 
\frac{\as}{2\pi} \;\left[ \frac{1}{1-\ga_N(\as)} + 
\frac{3}{2} \;\ga_N(\as) \right] \;
           + {\cal O}\left( \as^2(\as/N)^k \right) \;\;,
\end{equation}
\begin{eqnarray}\label{g22r}
\G_{22, \,N}(\as) = \frac{\as}{2\pi} \!\!\!&&\!\!\!\! \left\{ 
              \left( \frac{C_F}{C_A}
              \;C_{L, \,N}^{g \,(0)} - C_{L, \,N}^{S \,(0)} \right)
            \left[ \frac{1}{1-\ga_N(\as)} + \frac{3}{2} \;\ga_N(\as) \right] 
             \right. \nonumber \\
          \!\!\!&&\!\!\!\! \left. +  \left( \ga_{SS, \,N}^{(0)}
            - \frac{C_F}{C_A} \;\ga_{Sg, \,N}^{(0)} \right) 
            \right\} \;
           + {\cal O}\left( \as^2(\as/N)^k \right) \;\;.
\end{eqnarray}
Equations (\ref{g2lr},\ref{g22r}) provide resummed analytical formulae for
$\G_{2L}$ and $\G_{22}$ in terms of the BFKL anomalous dimension in 
Eq.~(\ref{gad}). Note that, if one compares the right-hand sides of these 
equations with the corresponding expressions for $\ga_{SS}$,
$\ga_{Sg}$, $C_L^a$ in Ref.~[\ref{CH}], one can see that 
Eqs.~(\ref{g2lr},\ref{g22r}) are remarkably simpler. These equations have to be
considered as the main scheme-invariant output of the next-to-leading order
resummation in the quark channel.

Having presented the main features of the physical anomalous dimensions 
$\G_{ij}$ both in fixed-order and in resummed perturbation theory, let me add
some comments on the dynamical equations (\ref{dF2p},\ref{dFLp}).

The first comment regards the theoretical accuracy at small $x$. Suppose, 
for instance, that the physical anomalous dimensions $\Gamma_{ij, \,N}(\as)$
are evaluated only to LL order in resummed perturbation theory. This implies
the following theoretical indeterminacy $\delta\Gamma_N/\Gamma_N$ 
$= {\cal O}(\as(\as/N)^k)$
or, equivalently, ${\cal O}(\as^2 (\as \ln x)^k)$ in $x$ space. In order to make
a direct comparison with the discussion in Sect.~\ref{fsd}, we can parametrize
this uncertainty in terms of a singular function of the type 
$A \as^2 \,x^{-K\as}$. Owing to the convolution structure in 
Eqs.~(\ref{dF2p},\ref{dFLp}) and considering the extreme case of flat structure
functions, this leads to the following theoretical uncertainty
\begin{equation}
\label{delF}
\frac{\delta F_{i=2,L}(x)}{F_{i=2,L}(x)} \sim \frac{A}{K} 
\; \as \;x^{-K\as} \;\;.
\end{equation}
Comparing Eqs.~(\ref{delf}) and (\ref{delF}), we can see that the replacement of
unphysical parton densities with physical observables (and the ensuing 
elimination of the factorization-scheme dependence) allows one to gain a factor
of $\as$ in the nominal theoretical accuracy. Of course, this is due to the fact
that the dynamical evolution equations (\ref{dF2p},\ref{dFLp}) to LL accuracy
contain more theoretical information than Eqs.~(\ref{F2}-\ref{dfg}) to the 
same accuracy. As a matter of fact, the evaluation of the physical anomalous
dimensions to LL order is equivalent to the knowledge of leading-order splitting
functions and next-to-leading order coefficient functions (see the discussion
above Eq.~(\ref{gl2r})).

Other comments regard phenomenological aspects. The evolution equation 
(\ref{dFLp}) for $F_L$ is physically analogous to the evolution equation 
for the gluon density. This analogy is particularly clear at small $x$, because
the physical anomalous dimensions $\Gamma_{LL}$ and $\Gamma_{L2}$ turn out
to be proportional to the BFKL anomalous dimension. Thus the effects of 
small-$x$ resummation in Eq.~(\ref{dFLp}) can directly be inferred from those 
studied in Ref.~[\ref{EHW}] for the gluon density.

The evolution equation 
(\ref{dF2p}) for $F_2$ is physically analogous to the evolution equation 
for the quark density. From the expressions in Eqs.~(\ref{g2lr},\ref{g22r})
we see that the small-$x$ resummation effects increase the amount of scaling
violation. Equation (\ref{g2lr}), for instance, can be rewritten as follows
\begin{equation}
\label{series}
\G_{2L, \,N}(\as) = 1 + 2.5 \;\ga_N(\as) + \sum_{n=2}^{+\infty} 
\left( \ga_N(\as) \right)^n \;\;. 
\end{equation}
Thus, besides the resummation accomplished by the BFKL anomalous dimension, 
there are further enhancing effects due to the positive definite (although, not
large) coefficients in the series (\ref{series}). Phenomenological studies
of these purely perturbative (i.e. independent of the parton densities)
effects appear interesting.

In general Eqs.~(\ref{dF2p},\ref{dFLp}) relate measurable values of observables,
$F_2$, $F_L$ and their derivatives with respect to $Q^2$, to perturbative
quantities, the physical anomalous dimensions. Thus, in the hard scattering 
regime, these equations provide absolute predictions of perturbative QCD.
In practice, the measurement of $dF_L/d\ln Q^2$ can be quite difficult. In this 
respect, once $F_L$ is measured at a certain value of $Q^2$, from 
Eq.~(\ref{dFLp}) one can obtain its value at any $Q^2$ and then one can use  
Eq.~(\ref{dF2p}) as a test of perturbative QCD that is free from 
non-perturbative parameters. Phenomenological analyses along these lines are
in progress [\ref{Thorne}].

\setcounter{footnote}{0}

\section{Parton picture and the unphysical gluon density}
\label{parton}

The mathematical steps that are necessary to go from 
Eqs.~(\ref{F2},\ref{dF2},\ref{dfg},\ref{FL}) to Eqs.~(\ref{dF2p},\ref{dFLp}) 
are pretty 
straightforward and, naively, one would be led to conclude that the former
equations are in one-to-one correspondence with the latter. This is not 
the case. In order to clarify this point let me first consider a case 
in which a one-to-one correspondence between physical and partonic observables
can really be established.

Suppose we want to evaluate the high-$Q^2$ behaviour of a hadronic observable 
$F_C$ other than, say, $F_2$ and $F_L$. Suppose also that it is a 
flavour-singlet observable measured in lepton-hadron scattering processes. Thus,
within the partonic framework, we should consider a factorization formula 
analogous to Eq.~(\ref{FL}). Writing this formula directly in $N$-space, we have
\begin{equation}\label{FCn}
F_{C, \,N}(Q^2) =  C_{C, \,N}^{S}(\as(Q^2)) \;{\tilde f}_{S, \,N}(Q^2)
+ C_{C, \,N}^{g}(\as(Q^2)) \;{\tilde f}_{g, \,N}(Q^2) \;\;. 
\end{equation}
Using the parton densities as determined from the scaling violations of 
(for instance) $F_2$ and $F_L$, the perturbative QCD prediction for $F_C$
in Eq.~(\ref{FCn}) amounts to the computation of {\em two} 
factorization-scheme dependent quantities: the coefficients functions 
$C_{C, \,N}^{S}(\as)$ and $C_{C, \,N}^{g}(\as)$.

Alternatively, we can use Eqs.~(\ref{F2},\ref{FL}), 
or Eqs.~(\ref{F2},\ref{dF2}), to rewrite Eq.~(\ref{FCn}) as follows
\begin{equation}
\label{FCk}
F_{C, \,N}(Q^2) =  K_{C2, \,N}(\as(Q^2)) \;F_{2, \,N}(Q^2)
+ K_{CL, \,N}(\as(Q^2)) \;F_{L, \,N}(Q^2) \;\;, 
\end{equation}
\begin{equation}
\label{FCkbar}
F_{C, \,N}(Q^2) =  {\overline K}_{C2, \,N}(\as(Q^2)) \;F_{2, \,N}(Q^2) +
\frac{1}{\G_{2C, \,N}(\as(Q^2))} \;\frac{d\ln F_{2, \,N}(Q^2)}{d\ln Q^2} \;\;, 
\end{equation}
where, using the same notation as in Eqs.~(\ref{gll}-\ref{g2l}), we have:
\begin{equation}\label{kfac}
K_{C2, \,N} = \frac{1}{\eav} \;\left[  C_{C, \,N}^{S} 
 - \frac{C_{L, \,N}^S}{C_{L, \,N}^g} C_{C, \,N}^{g} \right]_{{\rm DIS}} \;\;,
\;\;\;\;\;K_{CL, \,N} = \frac{1}{\eav} 
\;\left[  \frac{C_{C, \,N}^g}{C_{L, \,N}^g} \right]_{{\rm DIS}} \;\;.
\end{equation}
or:
\begin{equation}\label{kbarfac}
{\overline K}_{C2, \,N} = \frac{1}{\eav} \;\left[  C_{C, \,N}^{S} -
\frac{\ga_{SS, \,N}}{\ga_{Sg, \,N}} C_{C, \,N}^{g} \right]_{{\rm DIS}} \;\;,
\;\;\;\;\;\G_{2C, \,N} = \eav
\;\left[ \frac{\ga_{Sg, \,N}}{C_{C, \,N}^{g}} \right]_{{\rm DIS}} \;\;, 
\end{equation}

Equations (\ref{FCk}) and (\ref{FCkbar}) are equivalent to Eq.~(\ref{FCn}). The 
only difference is that the factorization-scheme dependence embodied in the 
coefficient functions $C_C^S$ and $C_C^g$ (and in the parton densities) has
been explicitly eliminated by replacing the parton densities with physical 
quantities ($F_2$ and $F_L$ in Eq.~(\ref{FCk}) or $F_2$ and its $Q^2$-derivative
in Eq.~(\ref{FCkbar})) and introducing the $K$-factors $K_{C2}, \,K_{CL}$ or
${\overline K}_{C2}, \,\G_{2C}$. These $K$-factors are factorization-scheme 
independent and have the same perturbative properties of
the physical anomalous dimensions (one of them, $\G_{2C}$, actually coincides 
with a physical anomalous dimension). As for the study of small-$x$ physics
and the comparison between fixed-order and resummed perturbation theory, 
the $K$-factors  are certainly preferred
[\ref{CCH},\ref{CAMICI}] with respect to the coefficient functions in 
Eq.~(\ref{FCn}). However, the perturbative QCD prediction for $F_C$ in 
Eq.~(\ref{FCk}) or (\ref{FCkbar}) still involves the computation of {\em two}
$K$-factors that replace the two coefficient functions.

The counting of `degrees of freedom' is instead different in the case of 
scaling violations. In order to describe the scaling violations of $F_2$ and
$F_L$ according to the partonic formulae in 
Eqs.~(\ref{F2},\ref{dF2},\ref{dfg},\ref{FL}), one has to assign two input
parton densities ${\tilde f}_S(x,Q_0^2), \,{\tilde f}_g(x,Q_0^2)$ (related
to the low-$Q^2$ behaviour of $F_2$ and $F_L$ or of $F_2$ and its $Q^2$-slope)
and to compute {\em six}\footnote{This number becomes eight in factorization 
schemes (like the \ms\ scheme) that are different from the DIS scheme.} 
quantities in QCD perturbation theory: four flavour-singlet splitting functions 
$P_{ab}(\as,x)$ and two coefficient functions $C_L^S(\as,x), \,C_L^g(\as,x)$.
On the contrary, the solution of the dynamical evolution equations
(\ref{dF2p},\ref{dFLp}) requires two non-perturbative initial conditions at the 
input scale $Q_0^2$ and the calculation of only {\em four} 
quantities, the physical anomalous dimensions $\G_{ij}(\as,x)$, 
in QCD perturbation 
theory. It is evident that the parton picture in 
Eqs.~(\ref{F2},\ref{dF2},\ref{dfg},\ref{FL}) introduces spurious perturbative 
QCD effects.  

Obviously, there is nothing wrong with the parton model or with the (light-cone)
Wilson 
expansion for deep-inelastic lepton-hadron scattering. The spurious effects
noticed above simply follow from the ambiguity in the definition of 
singlet-quark and gluon densities. From a field theory viewpoint, the ambiguity 
is related to the mixing under renormalization of singlet-quark and gluon
operators. Owing to the mixing matrix, the renormalization prescription
has to be specified by four (two, in DIS-type schemes)
arbitrary perturbative functions. In the partonic
framework this ambiguity is unavoidable and ultimately related to the fact that 
{\em no physical current} with point-like coupling to gluons does exist.

The unphysical perturbative contributions that are introduced through the 
definition of the gluon density are responsible for the theoretical 
uncertainty pointed out in Sect.~\ref{fsd}. In order to furtherly clarify this 
aspect, let me discuss another possible effect in resummed perturbation theory.

Suppose that in a certain factorization scheme the resummation of non-leading
logarithmic contributions in the quark or gluon anomalous dimensions produces
a singularity in the $N$-plane at a positive value $N={\overline N}(\as)= c_k
\as^k + \dots$. Solving the evolution equation (\ref{dfgn}) from the input
scale $Q_0^2$ to the hard scale $Q^2$, these higher-order terms factorize into 
$Q_0^2$-dependent and  $Q^2$-dependent contributions. In particular, the 
$Q_0^2$-dependent factor will contain the singularity at 
$N={\overline N}(\as(Q_0^2))$ and this singularity will dominate the small-$x$
behaviour of $F_2(x,Q^2)$ at {\em any} $Q^2$ unless it is cancelled by a zero
in the input parton densities ${\tilde f}_{a, \,N}(Q_0^2)$. In this case, 
however,
since the parton densities are assumed to be positive definite, they must have
a singularity at a value of $N$ larger than  ${\overline N}(\as(Q_0^2))$.
As a result the small-$x$ behaviour of $F_2(x,Q^2)$ turns out to be controlled
by an unphysical singularity that is simply due to the choice of the 
factorization scheme.

This is certainly an extreme effect. Nonetheless, it shows that the spurious 
perturbative functions that are introduced in the partonic picture may lead to 
{\em obstructions} that cannot any longer be removed within the same framework 
(i.e. without releasing the positivity constraint on the parton densities).

\section{Heavy-flavour structure functions}
\label{qqsec}

Most of the discussion in Sect.~\ref{inva} 
on physical anomalous dimensions can be repeated for 
other structure functions in deep-inelastic lepton-proton 
scattering, for instance, the heavy-flavour structure functions 
$F_2^{Q{\bar Q}}$ or $F_L^{Q{\bar Q}}$. 
These structure functions 
are completely analogous to the customary structure functions $F_2, \,F_L$
with the only additional constraint that heavy quarks of mass $M$ are produced
in the final state.

To be precise, all the theoretical formulae
in the previous Sections refer to the kinematical region $Q^2 \gg M^2$ and thus
neglect corrections of relative order $M^2/Q^2$. In order to take into account
the mass effects, 
one should perform the replacement $F_i \to F_i + F_i^{Q{\bar Q}}$. 
However, in the following $F_i$ still denotes the massless contribution to the 
structure function.

There are advantages and disadvantages in substituting the heavy-quark structure 
functions for $F_L$ in the study of scaling violations.
On the experimetal side [\ref{daum}], the charm contribution 
$F_2^{c{\bar c}}$ to the small-$x$ behaviour of the proton structure function 
is certainly more easily measurable than $F_L$.
This feature has to be contrasted with a larger theoretical uncertainty 
[\ref{vogt}] due to the unknown precise value of the charm mass and
with related complications considered below.

In the heavy-flavour case, the analogue of the collinear-factorization 
formula (\ref{FL}) is $(i=2,L)$:
\begin{eqnarray}
\label{facqq}
F_i^{Q{\bar Q}}(\xi,Q^2;M^2) &=&  \int_{\xi}^1 \frac{dz}{z} 
\left[ C_i^{Q{\bar Q}, \,g}(\as(Q^2),\xi/z;Q^2/M^2) 
\;{\tilde f}_g(z,Q^2) \right. \nonumber \\
&+& \left. C_i^{Q{\bar Q}, \,S}(\as(Q^2),\xi/z;Q^2/M^2) 
\;{\tilde f}_S(z,Q^2) \right] \;\; .
%+ \; \dots \; + {\cal O}(1/Q^2)  \;\; , 
\end{eqnarray}
Note that $F_i^{Q{\bar Q}}$ and the coefficient functions $C_i^{Q{\bar Q}, \,a}$
depend on the mass $M$. Also note that  
in Eq.~(\ref{facqq}) I have defined $F_i^{Q{\bar Q}}$ as function
of $Q^2, M^2$ and the {\em inelasticity variable} $\xi$, which is related 
to the customary Bjorken variable $x$ by $\xi=x(1+4M^2/Q^2)$. From a theoretical 
viewpoint the scaling variable $\xi$ is preferred to $x$ because it fulfils
the kinematical constraint $0 \leq \xi \leq 1$. Thus, considering  
$N$-moments with respect to $\xi$, Eq.~(\ref{facqq}) is diagonalized as
follows
\begin{equation}
\label{facqqn}
F_{i, \,N}^{Q{\bar Q}}(Q^2;M^2) =   
C_{i, \,N}^{Q{\bar Q}, \,g}(\as(Q^2);Q^2/M^2) 
\;{\tilde f}_{g, \,N}(Q^2) + C_{i, \,N}^{Q{\bar Q}, \,S}(\as(Q^2);Q^2/M^2) 
\;{\tilde f}_{S, \,N}(Q^2) \;\;. 
\end{equation}

The functions $C_i^{Q{\bar Q}, \,a}$  have a perturbative expansion similar
to Eq.~(\ref{cpert}). There are two main differences with respect to the case
of $C_L^{a}$. For fixed $\as$ the heavy-quark coefficient functions are not 
scale invariant because of their explicit dependence on $Q^2/M^2$. The expansion
for $C_{i, \,N}^{Q{\bar Q}, \,S}(\as;Q^2/M^2)$ starts in 
${\cal O}(\as^2)$ (i.e. $n \geq 2$) and thus only gluons contribute 
to Eq.~(\ref{facqq}) at LO (the sensitivity to the gluon density is somehow 
enhanced).

The coefficient functions $C_i^{Q{\bar Q}, \,a \,(n-1)}$
have been fully computed up to NLO $(n=0,1)$ in Ref.~[\ref{QQfo}].
The corresponding resummed formulae to NLL accuracy\footnote{The 
explicit formula for $C_{L, \,N}^{Q{\bar Q}, \,a}$ was not
reported in Ref.~[\ref{CCH}] and can be found in [\ref{DIS96}].}
were obtained in Ref.~[\ref{CCH}].

The evolution equations that involve the physical anomalous dimensions 
$\G^{Q{\bar Q}}$ for the pair of observables $\{F_2 , F_i^{Q{\bar Q}} \}$
are the following
\begin{eqnarray} 
\frac{dF_2(x,Q^2)}{d\ln Q^2} &=& \int^{1}_{x} \frac{dz}{z} \left[ 
\,\G_{22}^{Q{\bar Q}}(\as(Q^2),x/z;Q^2/M^2) \;
F_2(z, Q^2)  
\right. \nonumber \\ &+& \left. 
\G_{2i}^{Q{\bar Q}}(\as(Q^2),x/z;Q^2/M^2) 
\;F_i^{Q{\bar Q}}(z,Q^2;M^2) \right] \;\;,
\label{dF2pq} \\ 
%&+& 
%+ \; \dots \; + {\cal O}(1/Q^2)  
%\;\; , \nonumber
%\label{dF2pq} \\
\frac{dF_i^{Q{\bar Q}}(\xi,Q^2;M^2)}{d\ln Q^2} &=& \int^{1}_{\xi} 
\frac{dz}{z} \left[ 
\,\G_{i2}^{Q{\bar Q}}(\as(Q^2),\xi/z;Q^2/M^2) \;
F_2(z, Q^2)  
\right. \nonumber \\ &+& \left. 
\G_{ii}^{Q{\bar Q}}(\as(Q^2),\xi/z;Q^2/M^2) 
\;F_i^{Q{\bar Q}}(z,Q^2;M^2) \right] \;\; ,
%\nonumber \\ &+& 
%+ \; \dots \; + {\cal O}(1/Q^2)  \;\; ,\nonumber 
\label{dFipq} 
\end{eqnarray} 
or, equivalently, in $N$-space:
\begin{eqnarray}  
\frac{dF_{2, \,N}(Q^2)}{d\ln Q^2} &=& 
\G_{22, \,N}^{Q{\bar Q}}(\as(Q^2);Q^2/M^2) \;
F_{2, \,N}(Q^2) \nonumber \\
&+& \G_{2i, \,N}^{Q{\bar Q}}(\as(Q^2);Q^2/M^2) 
\;F_{i, \,N}^{Q{\bar Q}}(Q^2;M^2) \;, 
\label{dF2pqn} \\  
\frac{dF_{i, \,N}^{Q{\bar Q}}(Q^2;M^2)}{d\ln Q^2} &=&  
\,\G_{i2, \,N}^{Q{\bar Q}}(\as(Q^2);Q^2/M^2) \;
F_{2, \,N}(Q^2) \nonumber \\
&+& \G_{ii, \,N}^{Q{\bar Q}}(\as(Q^2);Q^2/M^2) 
\;F_{i, \,N}^{Q{\bar Q}}(Q^2;M^2) \;.  
\label{dFipqn}
\end{eqnarray} 

The relation between the physical anomalous dimensions 
$\G^{Q{\bar Q}}$ and the customary splitting and coefficient functions is 
similar to that in Eqs.~(\ref{gll}-\ref{g22}) for the case of 
$\{F_2 , F_L \}$, apart from the replacement
$C_{L, \,N}^a \to C_{i, \,N}^{Q{\bar Q}, \,a}/\eav$. Using the same notation 
as in Eqs.~(\ref{gll}-\ref{g22}), we have:
\begin{equation}\label{gii}
\G_{ii, \,N}^{Q{\bar Q}} = \left[ \ga_{gg, \,N} 
+ \frac{C_{i, \,N}^{Q{\bar Q}, \,S}}{C_{i, \,N}^{Q{\bar Q}, \,g}} 
 \;\ga_{Sg, \,N}
                    + \frac{d\ln C_{i, \,N}^{Q{\bar Q}, \,g}}{d\ln Q^2} 
                    \right]_{{\rm DIS}} \;\;,
\end{equation}
\begin{eqnarray}\label{gi2}
\G_{i2, \,N}^{Q{\bar Q}} &=& \frac{1}{\eav}
\left[ C_{i, \,N}^{Q{\bar Q}, \,g} \ga_{gq, \,N} 
     - C_{i, \,N}^{Q{\bar Q}, \,S} \ga_{gg, \,N}
            + C_{i, \,N}^{Q{\bar Q}, \,S}  \left( \ga_{SS, \,N} -
\frac{C_{i, \,N}^{Q{\bar Q}, \,S}}{C_{i, \,N}^{Q{\bar Q}, \,g}} 
\;\ga_{Sg, \,N} \right)
             \right. \nonumber \\
               &~& \;\; \left.                       
 + \; C_{i, \,N}^{Q{\bar Q}, \,S} 
 \left( \frac{d\ln C_{i, \,N}^{Q{\bar Q}, \,S}}{d\ln Q^2}
  - \frac{d\ln C_{i, \,N}^{Q{\bar Q}, \,g}}{d\ln Q^2} \right)
                    \right]_{{\rm DIS}} \;\;,
\end{eqnarray}
\begin{equation}\label{g2i}
\G_{2i, \,N}^{Q{\bar Q}} = \; \eav \;
\left[ \frac{\ga_{Sg, \,N}}{C_{i, \,N}^{Q{\bar Q}, \,g}} 
                     \right]_{{\rm DIS}} \;\;,
\end{equation}
\begin{equation}\label{g22q}
\G_{22, \,N}^{Q{\bar Q}} = \left[ \ga_{SS, \,N} 
       - \frac{C_{i, \,N}^{Q{\bar Q}, \,S}}{C_{i, \,N}^{Q{\bar Q}, \,g}} 
       \ga_{Sg, \,N}
                    \right]_{{\rm DIS}} \;\;.
\end{equation}

\subsection{Perturbative features}
\label{pfqq}

The dynamical evolution equations (\ref{dF2pq},\ref{dFipq}) 
(or (\ref{dF2pqn}),(\ref{dFipqn})) are
analogous to Eqs.~(\ref{dF2p},\ref{dFLp}) (or (\ref{dF2pn}),(\ref{dFLpn})). 
Note, however,
a main and important difference: now the physical kernels  
$\G^{Q{\bar Q}}$ depend not only on $\as$ but also on $Q^2/M^2$. This 
dependence is due to the mass-dependence of the heavy-flavour coefficients 
functions $C_{i}^{Q{\bar Q}, \,a}$. In particular, using the identity:
\begin{eqnarray}
\frac{d\ln C_{i, \,N}^{Q{\bar Q}, \,a}(\as(Q^2);Q^2/M^2)}{d\ln Q^2}
&=& \frac{d\ln\as(Q^2)}{d\ln Q^2} \;
\frac{\partial \ln C_{i, \,N}^{Q{\bar Q}, \,a}(\as(Q^2);Q^2/M^2)}
{\partial\ln\as(Q^2)} \nonumber \\
&+&
\frac{\partial \ln C_{i, \,N}^{Q{\bar Q}, \,a}(\as(Q^2);Q^2/M^2)}
{\partial \ln Q^2} \;\;,
\end{eqnarray}
we can see that the perturbative expansions of the physical anomalous 
dimensions $\G_{ii}^{Q{\bar Q}}$ and $\G_{i2}^{Q{\bar Q}}$ are the following
\begin{eqnarray}\label{giip}
&& \G_{ii, \,N}^{Q{\bar Q}}(\as;Q^2/M^2) = 
\frac{d\ln C_{i, \,N}^{Q{\bar Q}, \,g(0)}(Q^2/M^2)}{d\ln Q^2}
                  + \sum_{n=1}^{+\infty} \left( \frac{\as}{2\pi} \right)^{n} 
                   \;\G_{ii, \,N}^{Q{\bar Q} \,(n-1)}(Q^2/M^2) \nonumber \\
&& = 
\frac{d\ln C_{i, \,N}^{Q{\bar Q}, \,g(0)}(Q^2/M^2)}{d\ln Q^2}
+ \frac{\as}{2\pi} \left[ \G_{ii, \,N}^{Q{\bar Q} \,(0)}(Q^2/M^2)
         + \frac{\as}{2\pi} \; \G_{ii, \,N}^{Q{\bar Q} \,(1)}(Q^2/M^2)
            + \dots \right] \;\;,
\end{eqnarray}
\begin{eqnarray}\label{gi2p}
\G_{i2, \,N}^{Q{\bar Q}}(\as;Q^2/M^2) &=&
\sum_{n=1}^{+\infty} \left( \frac{\as}{2\pi} \right)^{n+1} 
                   \;\G_{i2, \,N}^{Q{\bar Q} \,(n-1)}(Q^2/M^2) \nonumber \\
&=& \left( \frac{\as}{2\pi} \right)^2 \left[ 
                     \G_{i2, \,N}^{Q{\bar Q} \,(0)}(Q^2/M^2) 
                    + \frac{\as}{2\pi} 
                    \;\G_{i2, \,N}^{Q{\bar Q} \,(1)}(Q^2/M^2)
                    + \dots \right] \;\;.
\end{eqnarray}
Comparing Eqs.~(\ref{giip}) and (\ref{gllp}), we see that 
$\G_{ii}^{Q{\bar Q}}$ contains a lowest-order contribution (the first term 
on the right-hand side of Eq.~(\ref{giip})) that is absent in $\G_{LL}$.
This contribution leads to {\em kinematical} scaling violations, that is, to
scaling violations that are independent of the running of $\as(Q^2)$ and
are simply due to the production kinematics of the heavy-quark pair. Thus, as
for the `true' dynamical scaling violations, the perturbative QCD calculation
should consistently
consider the coefficients $\G^{Q{\bar Q} \,(0)}$ in Eqs.~(\ref{giip},\ref{gi2p}) 
as LO terms, $\G^{Q{\bar Q} \,(1)}$ as NLO terms and so forth.

Apart from the explicit $(Q^2/M^2)$-dependence of the coefficients 
$\G^{Q{\bar Q} \,(n-1)}(Q^2/M^2)$, the perturbative expansions of 
$\G_{2i}^{Q{\bar Q}}$ and $\G_{22}^{Q{\bar Q}}$ are completely analogous to 
those in Eqs.~(\ref{g2lp}) and (\ref{g22p}), respectively.

Owing to its kinematical origin, the first term on the right-hand side of 
Eq.~(\ref{giip}) can be eliminated from the physical anomalous dimensions. To
this purpouse it is sufficient to rescale $F_{i}^{Q{\bar Q}}$ in 
Eqs.~(\ref{dF2pqn},\ref{dFipqn}) by the (factorization-scheme independent)
coefficient $C_{i}^{Q{\bar Q}, \,g(0)}$ and, thus, to consider
$F_{2, \,N}$ and $F_{i, \,N}^{Q{\bar Q}}/C_{i, \,N}^{Q{\bar Q}, \,g(0)}$
as dynamical variables. Nonetheless, this rescaling is not sufficient to exactly
put the physical anomalous dimensions $\G^{Q{\bar Q}}$ on equal terms with
those in Eqs.~(\ref{gllp}-\ref{g22p}). The interplay between kinematical and
dynamical scaling violations in the heavy-flavour case cannot be avoided
beyond the LO. The NLO coefficients $\G_{ii, \,N}^{Q{\bar Q} \,(1)}$ and 
$\G_{i2, \,N}^{Q{\bar Q} \,(1)}$ will always depend on the 
{\em next-to-next-to-leading} order (NNLO) (!) coefficient
functions $C_{i}^{Q{\bar Q}, \,a(2)}$. Since these have not yet been 
computed, a fully consistent NLO study of the dynamical 
evolution equation 
%(\ref{dF2pq},
(\ref{dFipq}) is, strictly speaking, not feasible at present.

This discussion of the perturbative features of the physical anomalous 
dimensions $\G^{Q{\bar Q}}$ is not peculiar to the heavy-quark case. It applies
to the physical anomalous dimensions of any structure function that depends
on some other large-momentum scale besides $Q^2$.

\subsection{Small-$x$ resummation}

The power counting of the logarithmic behaviour of $\G^{Q{\bar Q}}$ at small $x$
is similar to that of the physical anomalous dimensions relating $F_2$ and
$F_L$.

The two entries $\G_{ii}^{Q{\bar Q}}$ and $\G_{i2}^{Q{\bar Q}}$ have LL 
contributions. These can be obtained by using Eqs.~(\ref{gii},\ref{gi2}) and the 
known resummed formulae for the heavy-quark coefficient functions [\ref{CCH}]
and the quark anomalous dimensions [\ref{CHLett}]. I find:
\begin{equation}\label{giir}
\G_{ii, \,N}^{Q{\bar Q}}(\as;Q^2/M^2)
= \ga_N(\as) + 
\frac{\partial \ln H^{(i)}(\ga_N(\as);Q^2/M^2)}{\partial\ln Q^2} 
+ {\cal O}\!\left( \as(\as/N)^k \right) \;\;,
\end{equation}
\begin{eqnarray}
\G_{i2, \,N}^{Q{\bar Q}}(\as;Q^2/M^2) &=& \frac{1}{\eav} \;
\frac{\as}{2\pi} \left\{ \frac{C_F}{C_A} 
           \;C_{i, \,N}^{Q{\bar Q}, \,g \,(0)}(Q^2/M^2)
  \left[ \frac{}{} \G_{ii, \,N}^{Q{\bar Q}}(\as;Q^2/M^2) \right. \right.
  \nonumber \\
&-& \left.  
\frac{d\ln C_{i, \,N}^{Q{\bar Q}, \,g \,(0)}(Q^2/M^2)}{d\ln Q^2}
\left. \frac{}{} \right]
+\; {\cal O}\!\left( \as(\as/N)^k \right) \right\} \;\;,
\label{gi2r}   
\end{eqnarray}
where the functions $H^{(i)}(\ga;Q^2/M^2)$ in Eq.~(\ref{giir}) are simply
proportional to the $K$-factors $K_N^{(i)}(Q^2/M^2)$ introduced in the second
paper of Ref.~[\ref{CCH}]. Their eplicit expressions are: 
\begin{eqnarray}
&&\!\!\!H^{(2)}(\ga;Q^2/M^2) = \left(\frac{Q^2}{4M^2}\right)^{1-\ga} 
\left\{ 2(1+\ga) \frac{M^2}{Q^2} + 
\left[ 2 + 3\ga -3\ga^2 - 2(1+\ga)\frac{M^2}{Q^2}\right]
\right. \nonumber \\
&&\!\!\!\cdot \left. 
\left(1+\frac{Q^2}{4M^2}\right)^{\ga-1} 
\;F(1-\ga,1/2;3/2;\frac{Q^2}{Q^2+4M^2}) \right\} \;\;,
\label{H2}
\end{eqnarray}
\begin{eqnarray}
\!\!\!\!\!\!\!\!\!\!\!\!\!\!&&\!\!\!\!\!H^{(L)}(\ga;Q^2/M^2) = 
\left(\frac{Q^2}{4M^2}\right)^{1-\ga}
\frac{4M^2}{Q^2 + 4M^2} 
\left\{ \left( 1-\ga + \frac{6M^2}{Q^2} \right) \right. \nonumber \\
\!\!\!\!\!\!\!\!\!\!\!\!\!\!&&\!\!\!\!\!+ \left. 
\left[ \ga(1-\ga)\frac{Q^2}{2M^2} - 2(1-\ga) - \frac{6M^2}{Q^2}\right]
\left(1+\frac{Q^2}{4M^2}\right)^{\ga-1} 
\;F(1-\ga,1/2;3/2;\frac{Q^2}{Q^2+4M^2}) \right\} ,
\label{HL}
\end{eqnarray}
where $F(a,b;c;z)$ is the hypergeometric function.

Owing to the dependence on $Q^2/M^2$, the LL behaviour of 
$\G_{ii}^{Q{\bar Q}}$ (unlike that of $\G_{LL}$ in Eq.~(\ref{gllr})) is not
simply given by the BFKL anomalous dimension $\ga_N(\as)$. The resummation 
of the LL terms in $\G_{ii}^{Q{\bar Q}}$ is achieved through the 
$(\as/N)$-dependence of $\ga_N(\as)$ and the $\ga$-dependence of the function
$H^{(i)}(\ga;Q^2/M^2)$ on the right-hand side of Eq.~(\ref{giir}).
The LL contributions to $\G_{i2}^{Q{\bar Q}}$ in Eq.~(\ref{gi2r}) are 
proportional to $\G_{ii}^{Q{\bar Q}}$ after subtraction of its lowest-order
kinematic contribution (cf. Eq.~(\ref{giip})).

The evaluation of $\G_{ii}^{Q{\bar Q}}$ and $\G_{i2}^{Q{\bar Q}}$ to NLL 
accuracy would require the calculation of the gluon anomalous dimensions
$\ga_{ga}$ to NLL order and that of the heavy-flavour coefficient functions
to NNLL order. The NNLL accuracy in $C_i^{Q{\bar Q} \,a}$ is demanded by the
interplay between kinematical and dynamical scaling violations, as discussed in 
Sect.~\ref{pfqq}.

The anomalous dimensions $\G_{2i}^{Q{\bar Q}}$ and $\G_{22}^{Q{\bar Q}}$
contain only NLL terms at small $x$. These are explicitly given by the
following expressions
\begin{eqnarray}
\!\!\!\!\!\frac{\as}{2\pi} \;\G_{2i, \,N}^{Q{\bar Q}}(\as;Q^2/M^2) 
\!\!\!&=&\!\!\! \frac{\as}{2\pi}
\frac{N_f T_R \eav}{e^2_Q} \left[ 2 + 3\ga_N(\as) -3\ga^2_N(\as) \right]
\nonumber \\
\!\!\!&\cdot&\!\!\! \frac{{\sqrt \pi} 
\;\,\G(1+\ga_N(\as))}{\G(1/2+\ga_N(\as)) \;
H^{(i)}(\ga_N(\as);Q^2/M^2)}
+ {\cal O}\!\left( \as^2(\as/N)^k \right) \;,
\label{g2ir}
\end{eqnarray}
\begin{eqnarray}
\G_{22, \,N}^{Q{\bar Q}}(\as;Q^2/M^2) &=& \frac{\as}{2\pi} \left\{
\frac{C_F}{C_A} \,\frac{C_{i, \,N}^{Q{\bar Q}, \,g \,(0)}(Q^2/M^2)}{\eav} \;
\G_{2i, \,N}^{Q{\bar Q}}(\as;Q^2/M^2) \right. \nonumber \\
&+& \left. \left( \ga_{SS, \,N}^{(0)} 
- \frac{C_F}{C_A} \;\ga_{Sg, \,N}^{(0)} \right) \right\} \;
           + {\cal O}\left( \as^2(\as/N)^k \right) \;\;,
\label{g22rq}   
\end{eqnarray}
where $e_Q$ is the heavy-quark electric charge and $\G(z)$ is the Euler 
$\G$-function. The resummation of the logarithmic contributions in 
Eq.~(\ref{g2ir}) is embodied in the 
$(\as/N)$-dependence of BFKL anomalous dimension $\ga_N(\as)$ and the 
$\ga_N$-dependence of the functions $H^{(i)}$, according to 
Eqs.~(\ref{H2},\ref{HL}). In Eq.~(\ref{g22rq}) the physical 
anomalous dimension $\G_{22}^{Q{\bar Q}}$ to NLL accuracy is expressed
in terms of $\G_{2i}^{Q{\bar Q}}$ through a relation that is analogous
to that between $\G_{22}$ and $\G_{2L}$ in Eqs.~(\ref{g2lr},\ref{g22r}).

\section{Summary and discussion}
\label{summa}

In this contribution I have discussed how the study of different observables 
can contribute to our understanding of the dynamics of high-energy hadronic
interactions in the hard-scattering regime. The main motivation for considering
different observables is that from the analysis of a single quantity is 
difficult to disentangle perturbative from non-perturbative QCD physics.  
Of course, we aim to describe both perturbative and non-perturbative physics
but keeping separate the two aspects can simplify theoretical and 
phenomenological investigations.

In Sect.~\ref{gluon} the interplay between perturbative and 
non-perturbative  dynamics has been pointed out in the context of
QCD analyses of the small-$x$ behaviour of the proton structure function 
$F_2(x,Q^2)$. The factorization theorem of mass singularities provides a 
representation of $F_2$ in terms of phenomenological parton densities
and perturbatively computable splitting and coefficient functions. As long as
the latter have well-behaved perturbative expansions, this representation is 
highly predictive. In the small-$x$ regime, however, higher perturbative orders
are strongly enhanced by logarithmic contributions so that, in principle,
resummation procedures are mandatory. Thus a physical issue arises: where is
the boundary between perturbative and non-perturbative phenomena in the hard 
scattering regime? It is quite difficult to tackle this issue by studying the 
small-$x$ behaviour of the sole $F_2$. Indeed, as discussed in Sects.~\ref{fo}
and \ref{rpt}, the small-$x$ rise of $F_2$ produced by resumming LL and NLL 
contributions in the perturbative expansion of the splitting functions is, in
many respects (and with the present theoretical and experimental accuracy),
indistinguishable from a similar rise due to steep parton densities whose
$Q^2$ evolution is performed according to NLO perturbation theory. This 
uncertainty is formally taken into account by the factorization-scheme 
dependence, as
discussed in Sect.~\ref{fsd}. Owing to this dependence, the gluon density may 
play the role of a hidden variable that, in the case of $F_2$, relates different
perturbative QCD approaches, namely resummed and fixed-order perturbation 
theory. 

A better theoretical control on perturbative physics can be achieved by 
exploiting the very physical content of the factorization theorem, that is, the 
universality (process independence) of the parton densities. This means that
the same parton densities and the same perturbative approach have to be used
to study the small-$x$ behaviour of different physical observables. Universality
is particularly evident in the framework of the physical anomalous 
dimensions introduced in Sect.~\ref{inva}. Here I have discussed in detail the 
case 
of $F_2$ and $F_L$ but the method is completely general.

For any given set ${\tilde f}_a$ of parton densities one should consider a
set of an equal number of hadronic observables $F_a$. Thus, one can work out
the factorization procedure in matrix form as follows
\begin{equation}
\label{genft}
F =  C \;{\tilde f} \;\;,
\end{equation}
where $C=C_{ab}$ is the coefficient function matrix and the simple product 
structure on the right-hand side is usually valid in $N$-moment space. Then, it 
is straightforward to derive the following evolution equations
\begin{equation}
\label{pev}
\frac{dF}{d\ln Q^2} = \Gamma \;F \;\;,
\end{equation}
where the matrix $\Gamma$ of physical anomalous dimensions for the given set of
hadronic observables is related to $C$ and to the customary matrix $\gamma$
of anomalous dimensions as follows
\begin{equation}
\label{padgen}
\Gamma = \frac{dC}{d\ln Q^2} \;C^{-1} \; + C \;\ga \;C^{-1} \;\;.
\end{equation}
While $C$ and $\gamma$ are separately factorization-scheme dependent, the 
physical anomalous dimensions (\ref{padgen}) are factorization-scheme invariant.
As any other infrared and collinear safe observable, they are perturbatively 
computable apart from corrections that are suppressed by some inverse power of 
$Q$ in the hard-scattering regime. 

In Sect.~\ref{pad} I have considered the physical anomalous dimensions relating
the singlet components of $F_2$ and $F_L$. These are the most important 
contributions at small $x$ but the physical anomalous dimensions matrix can
be introduced in any kinematic region of $x$. It is just sufficient to
start from Eq.~(\ref{genft}) by including flavour non-singlet parton densities
and hadronic observables. 

In the small-$x$ region the theoretical and phenomenological importance of the
evolution equations (\ref{pev}) follows from the fact that the small-$x$
perturbative dynamics is completely controlled by the physical anomalous
dimensions. No spurious perturbative effect (see the discussion in 
Sect.~\ref{parton}) and no subtle interplay between perturbative logarithms and 
steepness of
parton densities takes place in Eq.~(\ref{pev}). The physical anomalous
dimensions can be evaluated both in fixed-order perturbation theory and in 
resummed perturbation theory. For any given set of observables and kinematic 
region of $x$, one can thus compare the two approaches and study the theoretical
accuracy of the perturbative expansion. Having done that, one can go back to the
partonic picture of Eq.~(\ref{genft}) and investigate more safely the 
small-$x$ behaviour of the non-perturbative parton densities. 

In Sect.~\ref{bsx}, I have presented resummed expressions for the physical 
anomalous dimensions $\Gamma_{LL},$ $\Gamma_{L2},$ $\Gamma_{2L},$ $\Gamma_{22}$.
Perturbative calculations to NLL accuracy are available for other hadronic 
observables and, in particular, for the heavy-flavour 
structure functions $F_2^{Q{\bar Q}}$ and $F_L^{Q{\bar Q}}$ [\ref{CCH}]. 
In Sect.~\ref{qqsec} I have considered the corresponding physical anomalous 
dimensions $\G^{Q{\bar Q}}$.
Using the theoretical approach of Refs.~[\ref{CCH}-\ref{LRSS}]
one can investigate the small-$x$ behaviour of
other physical anomalous dimensions. In my opinion, some phenomenological
studies within the framework of physical anomalous dimensions are warranted. 

\vskip 0.8 true cm

\noindent {\bf Acknowledgments.  } The results reported in this paper were 
presented and discussed at the 
UK Phenomenology Workshop on HERA Physics, St. John's College, Durham, UK,
September 1995. I would like to thank Robin Devenish and Mike Whalley
for the excellent organization of the meeting. 
\newline
I also would like to thank Marcello Ciafaloni and Francesco Hautmann
for useful comments, discussions and a collaboration over many years
on these topics.

\vskip 0.5 true cm
\noindent {\bf Note added. } The use of physical anomalous dimensions has been
previously advocated in the literature (see, for instance, Ref.~[\ref{PAD}])
although in different context. I wish to thank Georges Grunberg for
having pointed out those references to me. 

\vskip 0.3 true cm
{\large \bf References}
\begin{enumerate}
%\normalsize

\item \label{HERA}
ZEUS Coll., M.\ Derrick et al., \pl{316}{412}{93}, \zp{65}{379}{95}, 
\zp{69}{607}{96}, preprint DESY 96-76;
H1 Coll., I.\ Abt et al., \np{407}{515}{93};
H1 Coll., T.\ Ahmed et al., \np{439}{471}{95}. 

\item \label{C} 
      S.\ Catani, in Proceedings of {\it Les Rencontres de Physique de 
      La Vall\'{e}e d'Aoste}, ed. M.\ Greco 
      (Editions Frontieres, Gif-sur-Yvette, 1994), p.~227.

\item \label{first}
      A.\ De R\'{u}jula, S.L. Glashow, H.D.\ Politzer, S.B.\ Treiman, 
      F.\ Wilczek and A.\ Zee, \pr{10}{1649}{74}.
      
\item \label{BFKL}
      L.N.\ Lipatov, Sov. J. Nucl. Phys. 23 (1976) 338; E.A.\ Kuraev,
      L.N.\ Lipatov and V.S.\ Fadin, Sov. Phys. JETP  45 (1977) 199; Ya.\
      Balitskii and L.N.\ Lipatov, Sov. J. Nucl. Phys. 28 (1978) 822.

\item \label{CCH}
      S. Catani, M. Ciafaloni and F. Hautmann, Phys. Lett. B242
      (1990) 97, Nucl. Phys. B366 (1991) 135, and 
      in Proceedings of the
      HERA Workshop, eds. W.\ Buchm\"{u}ller and G.\ Ingelman (DESY, Hamburg, 
      1991), p.~690.

\item \label{CE}
      J.C. Collins and R.K. Ellis, Nucl. Phys. B360 (1991) 3. 

\item \label{LRSS}
      E.M.\ Levin, M.G.\ Ryskin, Yu.M.\ Shabel'skii and A.G.\ Shuvaev, Sov. J.
      Nucl. Phys. 53 (1991) 657.

\item \label{CCHLett}
      S.\ Catani, M.\ Ciafaloni and F.\ Hautmann, \pl{307}{147}{93}.

\item \label{CHLett}
      S.\ Catani and F.\ Hautmann, \pl{315}{157}{93}.

\item \label{CH}
      S.\ Catani and F.\ Hautmann, \np{427}{475}{94}.

\item \label{AP}
       V.N.\ Gribov and L.N.\ Lipatov, Sov. J. Nucl. Phys. 15 (1972) 438, 
       675; G.\ Altarelli and G.\ Parisi,
       \np{126}{298}{77}; Yu.L.\ Dokshitzer, Sov. Phys. JETP  46 (1977) 641.

\item \label{CSS}
       See, for instance, J.C.\ Collins, D.E.\ Soper and G.\ Sterman, in
       {\it Perturbative Quantum Chromodynamics}, ed. A.H.\ Mueller
       (World Scientific, Singapore, 1989), p.~1 and references therein.

\item \label{DIS}
      G.\ Altarelli, R.K.\  Ellis and G.\ Martinelli, \np{157}{461}{79}.

\item \label{CFP}
      E.G.\ Floratos, D.A.\ Ross and C.T.\ Sachrajda, \np{152}{493}{79};
      A.\ Gonzalez-Arroyo, C.\ Lopez and F.J.\ Yndurain, \np{153}{161}{79};
      G.\ Curci, W.\ Furmanski and R.\ Petronzio,  \np{175}{27}{80};  
      W.\ Furmanski and R.\ Petronzio, \pl{97}{437}{80};
      A.\ Gonzalez-Arroyo and C.\ Lopez, \np{166}{429}{80};
      E.G.\ Floratos, C.\ Kounnas and R.\ Lacaze, \pl{98}{285}{81}.
%      \np{192}{417}{81}.

\item \label{fit}
%       H1 Coll., I. Abt et al., \pl{321}{161}{94};
       ZEUS Coll., M.\ Derrick et al., \pl{345}{576}{95}; 
       H1 Coll., S.\ Aid et al., \pl{354}{494}{95}, preprint DESY 96-39.

\item \label{GRV}
      M.\ Gl\"{u}ck, E.\ Reya and A.\ Vogt, \zp{67}{433}{95}. 

\item \label{CTEQ}
      CTEQ Coll., H.L.\ Lai et al., \pr{51}{4763}{95}, preprint MSUHEP-60426
      (hep-ph/9606399).

\item \label{MRS}
      A.D.\ Martin, R.G.\ Roberts and W.J. Stirling, \pr{50}{6734}{94},
      \pl{354}{155}{95}, preprint DTP/96/44 (hep-ph/9606345). 

\item \label{DOUBLE}
      R.D.\ Ball and S.\ Forte, 
      \pl{335}{77}{94}, %      double asymptotic scaling
      \pl{336}{77}{94}, %      LO analysis
      Nucl. Phys. B (Proc. Supp.) 
      39BC (1995) 25.    %      NLO analysis

\item \label{Ciaf}
       M.\ Cia\-fa\-lo\-ni,  \pl{356}{74}{95}.

\item \label{Jar}
       T. Jaroszewicz, Phys. Lett. B116 (1982) 291.

\item \label{FLF}
      L.N.\ Lipatov and V.S.\ Fadin, \sj{50}{712}{89}; V.S.\ Fadin and
      R.\ Fiore, \pl{294}{286}{92}; V.S.\ Fadin and L.N.\ Lipatov, 
      \np{406}{259}{93}; 
      V.S.\ Fadin, R.\ Fiore and A.\ Quartarolo, 
%      \pr{50}{2265}{94},
       \pr{50}{5893}{94}, \pr{53}{2729}{96}; 
      V.S.\ Fadin, Phys. Atom. Nucl. 58 (1995) 1762; 
      V.S.\ Fadin and L.N.\ Lipatov, preprint DESY-96-020 (hep-ph/9602287);
      V.S.\ Fadin, R.\ Fiore and M.I. Kotskii, preprint BUDKERINP-96-35
      (hep-ph/9605357). 

\item \label{DD}
      V.\ Del Duca, \pr{54}{989}{96}, preprint EDINBURGH 96-3 
      (hep-ph/9604250).

\item \label{Cam}
      G.\ Camici and M.\ Ciafaloni, preprint DFF 250-6-96 (hep-ph/9606427)
      and to appear.
      
\item \label{EHW}
       R.K.\ Ellis, F.\ Hautmann and B.R.\ Webber, \pl{348}{582}{95};
       F.\ Hautmann, in {\it QCD and High Energy Hadronic Interactions},
       Proceedings 30th Rencontres de Moriond, ed. J. Tran Thanh Van
       (Editions Frontieres, Gif-sur-Yvette, 1995), p.~133.

\item \label{BFres}
       R.D.\ Ball and S.\ Forte, \pl{351}{313}{95}, %      small-x summation
      \pl{358}{365}{95}. %      \as and fits 

\item \label{AKMS}
       A.J.\ Askew, J.\ Kwiecinski, A.D.\ Martin and P.J.\ Sutton, 
       \pr{47}{3775}{93}, \pr{49}{4402}{94}; A.\ J. Askew, 
       K.\ Golec Biernat, J.\ Kwiecinski, A.D.\ Martin 
       and P.J.\ Sutton, \pl{325}{212}{94}. 

\item \label{FRT}
       J.R.\ Forshaw, R.G.\ Roberts and R.S.\ Thorne,      
       \pl{356}{79}{95}.

\item \label{SDIS}
       S.\ Catani, \zp{70}{263}{96}.
       
\item \label{subeffects}
       S.A.\ Larin, P.\ Nogueira, T.\ van Ritbergen, J.A.M.\ Vermaseren, 
       preprint NIKHEF-96-010 (hep-ph/9605317); 
       S.\ Forte and R.D.\ Ball, preprint EDINBURGH-96-14 (hep-ph/9607291), 
       to appear in Proceedings
       of International Workshop on Deep Inelastic Scattering and
       Related Phenomena (DIS 96), Rome, Italy, April 1996;
       J.\ Bl\"{u}mlein, S.\ Riemersma and A. Vogt, preprint 
       HEPPH-9607329 (hep-ph/9607329), to appear in Proceedings
       of International Workshop on Deep Inelastic Scattering and
       Related Phenomena (DIS 96), Rome, Italy, April 1996.

\item \label{BFscheme}
       R.D.\ Ball and S.\ Forte, \pl{359}{362}{95}. %      mom cons and scheme 

\item \label{CAMICI}
      G.\ Camici and M.\ Ciafaloni, \np{467}{25}{96}.

\item \label{LCF}
       D.I.\ Kazakov, A.V.\ Kotikov, G.\ Parente, O.A.\ Sampayo
       and J.\ S\`{a}nchez-Guille\`{e}n, \prl{65}{1535}{90};
       J.\ S\`{a}nchez-Guille\`{e}n, J.L.\ Miramontes, M.\ Miramontes,
       G.\ Parente and O.A.\ Sampayo, \np{353}{337}{91};
       E.B.\ Zijlstra and W.L.\ van Neerven, \np{383}{525}{92}.
       
\item \label{Thorne}
       R.S.\ Thorne, 
       talk at International Workshop on Deep Inelastic Scattering and
       Related Phenomena (DIS 96), Rome, Italy, April 1996;
       R.G.\ Roberts and R.S.\ Thorne, in preparation.     

\item \label{daum}
       K.\ Daum, talk at International Workshop on Deep Inelastic Scattering and
       Related Phenomena (DIS 96), Rome, Italy, April 1996.

\item \label{vogt}
       M.\ Gl\"{u}ck, A. Reya and M.\ Stratmann, \np{422}{37}{94};
       A.\ Vogt, preprint DESY 96-012 (hep-ph/9601352).
       
\item \label{QQfo}
      E.\ Laenen, S.\ Riemersma, J.\ Smith and W.L.\ van Neerven,
      \np{392}{162}{93}; S.\ Riemersma, J.\ Smith and W.L.\ van Neerven,
      \pl{347}{143}{95}.

\item \label{DIS96}
      S.\ Catani, 
      preprint DFF 254-7-96 (hep-ph/9608310), to appear in Proceedings
      of International Workshop on Deep Inelastic Scattering and
      Related Phenomena (DIS 96), Rome, Italy, April 1996.

\item \label{PAD}
       W.A.\ Bardin and A.J.\ Buras, \pl{86}{61}{79}; E.G.\ Floratos,
       C.\ Kounnas and R.\ Lacaze, \np{192}{417}{81}; G.\ Grunberg,
       \pr{29}{2315}{84}.

\end{enumerate}

\end{document}